\documentclass[12pt]{iopart}

 \usepackage{iopams}
\usepackage{braket}
\usepackage{amsfonts}
\usepackage{color}
\usepackage[dvipsnames]{xcolor}
\usepackage{amssymb}
\usepackage{mathrsfs}
\usepackage{graphicx}
\usepackage{lmodern}
\usepackage{lineno}
\usepackage{hyperref}
\usepackage[normalem]{ulem}
\expandafter\let\csname equation*\endcsname\relax
\expandafter\let\csname endequation*\endcsname\relax
\usepackage{amsmath}
 \usepackage[mathscr]{euscript}
 \usepackage[all]{xy}
\newtheorem{definicion}{Definition}[section]

\newtheorem{observacion}[definicion]{Remark}

\newtheorem{notacion}[definicion]{Notation}

\newcommand{\beq}{\begin{eqnarray}}
\newcommand{\eeq}{\end{eqnarray}}

\def\keywords#1{\vspace{10pt}
     \begin{indented}
     \item[]\rm Keywords: #1\par
     \end{indented}}

\begin{document}



\title{Geometric formulation for Palatini-Cartan gravity }
\author{Jasel Berra-Montiel, Iván Cortes-Cruz, and  Alberto Molgado}
\address{Facultad de Ciencias, Universidad Aut\'onoma de San Luis 
Potos\'{\i} \\
Campus Pedregal, Av. Parque Chapultepec 1610, Col. Privadas del Pedregal, San
Luis Potos\'{\i}, SLP, 78217, Mexico.}

\eads{\mailto{\textcolor{blue}{jasel.berra@uaslp.mx}}, \mailto{\textcolor{blue}{A391738@alumnos.uaslp.mx}}, \mailto{\textcolor{blue}{alberto.molgado@uaslp.mx}}
}



\begin{abstract}   
Motivated by the increasing efforts to understand the covariant structure of physical models associated with General Relativity using different kinds of geometric frameworks, in this article we analyze the four-dimensional Palatini-Cartan model for gravity, which is a well-known generalization of General Relativity, from the perspective of various geometric-covariant formalisms for classical field theory. At the Lagrangian level, we do not only recover the correct field equations of the theory, which are equivalent to the torsion-free condition and the Einstein equations, but we also study the gauge symmetries of the model in order to construct the Lagrangian momentum map associated with the action of the gauge symmetry group on the configuration space of the system and, consequently, its corresponding Noether currents. Within the multisymplectic approach, we analyze the action of the gauge symmetry group on the multi-momenta phase space of the model,  and we also introduce the induced momentum map that allows us to recover the admissible Cauchy data of the system. Further, we also apply the algorithm to treat singular systems within the polysymplectic framework, in which, in order to obtain the correct field equations of the model, we introduce a non-trivial Dirac-Poisson bracket characterized by the generalized Moore-Penrose inverse of the matrix induced by the second class constraints of the system. Finally, using the multisymplectic framework as a starting point, we perform the space plus time decomposition of the system to recover the instantaneous Lagrangian and the extended Hamiltonian of the theory, as well as the gauge structure that characterize the Palatini-Cartan model for gravity within the instantaneous Dirac-Hamiltonian formalism.

\end{abstract}



\section{Introduction}
A well-known approach to the study of classical field theories is the instantaneous Dirac-Hamilton formalism~\cite{henb1}, which provides a good understanding of the gauge symmetries and the local degrees of freedom of a given classical field theory. However, since we must identify a temporal direction, the instantaneous Dirac-Hamiltonian formulation for a classical field theory starts from a foliation of the spacetime manifold into Cauchy surfaces which, consequently, obscures the true covariant nature of the system. As a result, the implementation of the canonical quantization scheme might lead to quantum field theories whose explicit covariance is no longer obvious~\cite{forger1}. In contrast, the multisymplectic formalism, which is closely related to the De Donder-Weyl canonical Hamiltonian theory~\cite{dedonder1}, provides a finite-dimensional, geometric, and covariant Hamiltonian-like formulation for classical field theories. The origins of the multisymplectic formalism are related to De Donder and Weyl~\cite{dedonder1, weyl1}. But its relationship with classical field theory was only developed at the end of the last century mainly by Polish school in the works of Kijowski, Szczyrba and Tulczyjew~\cite{ki1, sz1, tu1}. Further progress was introduced in related research~\cite{gotay1, gotay2, cari1, deleon2, molgado1}. On the one hand,  the multisymplectic formalism stars by identifying the fields of a theory as sections of an adequate fiber bundle and considering a multiphase space (or multi-momenta space), which is a finite-dimensional space defined locally by assigning to each coordinate $n$ conjugate momenta, where $n$ is the dimension of the underlying spacetime manifold. We can endow such a multiphase space with a multisymplectic $(n+1)$-form, which constitutes the main geometric object within the multisymplectic approach since it allows not only to obtain the correct field equations but, also, to describe the symmetries of the theory under study. In particular, the infinitesimal action of the symmetry group on the multiphase space induces the so-called covariant momentum map, which allows one to extend the Noether theorems to the multisymplictic framework and at the same time give rise to the momentum map that characterized the system from the instantaneous Hamiltonian point of view~\cite{gotay2}. Further, the above is important for classical field theories whose symmetries are localizable since, in light of the second Noether theorem, it results that the admissible Cauchy data for the evolution of the system coincide with the zero level set of such momentum map. Thus,  for such classical field theories, as it is explained in~\cite{gotay2}, the vanishing of the momentum map will give rise to the set of first-class constraints of the system, in Dirac's terminology. As a consequence, we can build a bridge between multisymplectic structures and the constrained structure of a given singular Lagrangian system. In addition, a natural question within the multisymplectic approach is related to the analogous of Poisson structures bearing a relation to multisymplectic forms that extends the way in which Poisson structures generalize symplectic structures. As an example, in the multimomenta phase space we can construct an analogous to the Poisson-bracket over the space of the so-called Hamiltonian $(n-1)$-forms but the Jacobi identity usually fails~\cite{ibort1, forger2}. A lot of work has been done to adress with this issue, however, most of such approaches involve a Lie super-algebra but not a Poisson super-algebra~\cite{hr1}. This is unsatisfactory from the point of view of physics because the introduction of a well-defined Poisson bracket represents the initial step towards either a pre-canonical quantization scheme or a deformation quantization associated with the multisymplectic formalism~\cite{kana5}.\\
On the other hand, within the polysymplectic formalism for classical field theories we can define a canonical $(n+1)$-form, called the polysymplectic form, on a subspace of the multimomenta phase space known as the polymomentum phase space. Such a $(n+1)$-form allow us to construct a well-defined generalized Poisson bracket, which is suitable for quantization purposes, and leads to a Poisson-Gerstenhaber algebra structure with respect to the graded Lie bracket and the co-exterior product of forms. In particular, with this Poisson structure at hand, we can analyze any arbitrary classical field theory in a canonical Poisson-Hamiltonian framework. For some examples see~\cite{molgado1, kanaex1, kana1, vey1, deleon1, molgadoex1, molgado2, molgadoex2}, and references therein.\\
However, the analysis of the action of the symmetry group for a given field theory  from the point of view of such geometric formulations has been rarely explored, specifically when we consider non-trivial models for General Relativity. Therefore, one of our objectives here is to apply this kind of formalism to describe the gauge structure of the Palatini-Cartan model for gravity. The reformulation of General Relativity as a gauge theory involves replacing the theory's metric approach with a formulation in terms of a dynamical connection, which is a section of a fiber bundle over spacetime, gauging a local structure. Now, within the Palatini formulation of General Relativity, this connection $\omega$ is a section of a Lorentz group principal bundle over the spacetime manifold $X$. However, an additional ingredient is needed, which has no analogous, for example, in the Yang-Mills theory, namely a co-frame field $e$ (sometimes called a soldering form or a tetrad in the context of 4-dimensional gravity). But, since the end of the last century, some people have started to consider the connection $\omega$ and the soldering form $e$ as part of a single field $A=\omega + e$, which can be shown to be a Cartan-connection~\cite{mm1, stells1, catren1}. The above allows us to view the Palatini model for gravity in terms of Cartan geometry~\cite{catren1, wise1, wise2}, so we shall refer to it hereafter as the Palatini-Cartan model for gravity. Bearing this in mind, our main objective is to analyze the Palatini-Cartan model for gravity from the point of view of the covariant-geometric approach for classical field theory. After briefly describe the mathematical tools to formulate the Palatini-Cartan model for gravity, we analyze it from a geometric-covariant Lagrangian approach. To be more specific, we obtain the field equations and analyze its gauge symmetries, which allow us to build the covariant Lagrangian momentum map induced by the action of the extended symmetry group of the theory and, consequently, its Noether currents. As a result, we prove that the symmetries of the theory correspond to localizable symmetries. Next, we discuss the Palatini-Cartan model for gravity within the multisymplectic formalism. After introducing the multisymplectic form associated with the theory, we focus our attention on describing how the action of the symmetry group gives rise to canonical transformations on the multimomenta phase space, which also allow us to construct a momentum map associated with the action of the symmetry group of the theory. Further, in order to perform a consistent polysymplectic formulation for the Palatini-Cartan model for gravity, we follow the algorithm proposed in~\cite{kana4} to analyze singular Lagrangian systems within the polysymplectic approach. In Dirac's terminology, we find that within the polysymplectic approach, the system is characterized by a set of second-class constraints $(n-1)$-forms. To identify this second class constraints as strong identities, we need to construct a Dirac-Poisson bracket in order to obtain the correct De Dorder-Weyl-Hamiltonian field equations of the theory, that is, the torsion free condition and the Einstein field equations. We would like to point out that, due to the structure of the matrix defined by the sencond-class contraints of the model, we must introduce a generalized Moore-Penrose inverse in order to construct a well-defined Dirac-Poisson bracket, as proposed in~\cite{kana1}, but here we provide the explicit form of such a Moore-Penrose generalized inverse. Furthermore, in~\cite{molgado1} was a non-trivial Dirac-Poisson bracket for a three-dimensional model for gravity introduced, which is a purely topological model without local degrees of freedom, however, as far as we know, here is the first time that a non-trivial Dirac–Poisson bracket has been explicitly constructed within a covariant geometric framework for a four-dimensional gravitational model that is both dynamical and propagates two local degrees of freedom, thereby providing a realistic and non-trivial description of gravity. A well-defined Dirac-Poisson bracket is also relevant at the quantum level, since, within the polysymplectic formulation, it constitutes the first step toward a precanonical quantization of General Relativity based in the De Donder-Weyl-Hamiltonian formalism~\cite{kana5}. In addition, we briefly discuss the way in which the momentum map associated with the extended gauge symmetry group of the system allows us to obtain the conserved currents of the theory within the De Donder-Weyl canonical formalism.  Finally, following~\cite{gotay1, gotay2, deleon1}, we perform the space plus time decomposition of the Palatini-Cartan model for gravity, within both Lagrangian and multisymplectic levels. This allows us to recover not only the instantaneous extended Hamiltonian, but also the set of first- and second-class constraints that characterize the system within the instantaneous Dirac-Hamiltonian formulation, as it is described in~\cite{nikolic1}.\\
This paper is organized as follows: in section~\ref{sec2}, we review the Lagrangian, multisymplectic and polysymplectic formalisms for classical field theory. In addition, we briefly describe the space plus time decomposition for a given classical field theory, which results to be fundamental to stablish a relation between the multisymplectic and the instantaneous Dirac-Hamiltonian formalisms. In section~\ref{sec3}, we formally define the Palatini-Cartan model for gravity. Next, we proceed to analyze it within the geometric-covariant Lagrangian, multisymplectic and polysymplectic formulations for classical field theory, focusing in the study of both the field equations and the gauge symmetries of the model. Further, we perform the space plus time decomposition of the theory both at the Lagrangian and the multisymplectic level, which allows us to recover the instantaneous Dirac-Hamiltonian description of the theory. Finally, section~\ref{sec4} is devoted to some concluding remarks and commentaries.

\section{Geometric-covariant analysis for classical field theories. } \label{sec2}
In this section, we briefly review the geometric-covariant formalism for classical field theories closely following~\cite{gotay1, gotay2, deleon1, kana0, kana2, kana3}. \\
We start by introducing some conventions used throughout this paper. To start with, let $M$ be a manifold, then we can define a fiber bundle as follows.
\begin{definicion}
    The triple $(E, \pi_{ME}, M)$ is called a fiber bundle, where $M$ stands for the base manifold, $E$ is the total space (a fiber manifold), and $\pi_{ME}: E \rightarrow M$ represents the projector map. At $p\in M$, the subset $E_{p}:= \pi_{ME}^{-1}(p) $ is called the fiber of $E$ over $p$. 
\end{definicion}
As usual, $\mathscr{E}_{M}:=\{\phi:M \rightarrow E \left. | \right. \pi_{ME} \circ \phi = id_{M} \}$ denotes the set of sections of $\pi_{ME}$\footnote{We will henceforth refer to a fiber bundle $(E, \pi_{ME}, M)$ simply as $\pi_{ME}$.}. Moreover, given the tangent bundle of $M$, $(TM, \pi_{MTM}, M)$, and its cotangent bundle, $(T^{\ast}M, \pi_{MT^{\ast}M}, M)$, we denote their $k$-th exterior product bundle as $(\bigwedge^{k} TM, \pi_{M\bigwedge^{k} TM}, M)$ and $(\bigwedge^{k} T^{\ast}M, \pi_{M\bigwedge^{k}T^{\ast}M}, M)$, respectively.
\begin{notacion}
    The sets $\mathfrak{X}^{k}(E)$ and $\Omega^{k}(E)$ denote the set of sections of $\pi_{M\bigwedge^{k} TM}$ and $\pi_{M\bigwedge^{k}T^{\ast}M}$, respectively. 
\end{notacion}
In addition, given a fiber bundle $\pi_{ME}$, we define $(VE, \pi_{EVE},E)$ as its vertical tangent bundle, whose fibers over $q\in E$ are given by 
\begin{equation}
    V_{q}E:= \{v \in T_{q}E \left. | \right. d\pi_{ME}{}_{q}(v)=0 \in T_{\pi_{ME}(q)}M \}
\end{equation}
being $d\pi_{ME}{}_{q}: T_{q}E \rightarrow  T_{\pi_{ME}(q)}M$ the tangent map of $\pi_{ME}$ at $q\in E$.  
\begin{definicion}
    We define $(\bigwedge^{k}_{r}T^{\ast}E, \pi_{E\bigwedge^{k}_{r}T^{\ast}E}, E)$ as the vector bundle of horizontal $(k, r)$-forms over $E$, whose fiber over $q \in E$ is given by 
    \begin{equation}
        \bigwedge{}^{k}_{r}T^{\ast}E:=\{ \omega \in \bigwedge{}^{k}T^{\ast}E \left. | \right. v_{1}\lrcorner \cdots v_{r+1}\lrcorner \omega=0,\, \forall v_{1},\cdots, v_{r+1} \in V_{q}E\}
    \end{equation}
\end{definicion}
Hence, we use $\Omega_{r}^{k}(E)$ to denote the set of sections of $\pi_{E\bigwedge^{k}_{r}T^{\ast}E}$. \\

Now, in the context of classical field theory, given a spacetime manifold $X$ whose local coordinates are $(x^{\mu})$, the configuration space of the system can be defined by a fiber bundle $(Y, \pi_{XY}, X)$, whose fiber coordinates will be denoted by $(y^{a})$. An adapted coordinate system in $Y$ is $(x^{\mu}, y^{a})$. The dynamical fields can be identified with sections $\phi \in \mathscr{Y}_{X}$ of the bundle $\pi_{XY}$, whose local representation is given as usual by $(x^{\mu}, \phi^{a})$, where $\phi^{a}:= y^{a}\circ \phi$. Therefore, it is natural that the Lagrangian dynamics takes place in the first order jet bundle $(J^{1}Y, \pi_{XJ^{1}Y}, X)$. To define it, we need to introduce the affine jet bundle  $(J^{1}Y, \pi_{YJ^{1}Y}, Y)$, whose fibers at $y \in Y$ are given by $J_{y}^{1}Y:= \{ \gamma \in L(T_{x}X, T_{y}Y) | d\pi_{XY} \circ \gamma = id_{T_{x}X}\}$, where $L(T_{x}X, T_{y}Y)$ is the set of linear maps from $T_{x}X$ to $T_{y}Y$, with $x:= \pi_{XY}(y)$. Note that, the affine space $J_{y}^{1}Y$ is modeled over the vector space $L(T_{x}X, V_{y}Y)$, where $V_{y}Y$ is the vertical tangent space of $Y$ at $y$. It follows that an adapted coordinate system on $J_{y}^{1}Y$ is defined by $(x^{\mu}, y^{a}, y_{\mu}^{a})$, where $y_{\mu}^{a}$ denotes the so-called derivative coordinates. Thus, the jet bundle $\pi_{XJ^{1}Y}$ is given by $\pi_{XJ^{1}Y}:= \pi_{XY} \circ \pi_{YJ^{1}Y}$. In particular, given a section $\phi \in \mathscr{Y}_{X}$ we can induce a section $j^{1}\phi \in \mathscr{J}^{1}{}\mathscr{Y}_{X}$, called the first jet prolongation of $\phi$, whose coordinate representation is denoted by $(x^{\mu}, \phi^{a}, \phi_{\mu}^{a})$, being $\phi_{\mu}^{a}:= \partial_{\mu}\phi^{a}$. For more details on jet bundles, see~\cite{book1, book2}.
\subsection{Geometric-covariant Lagrangian formalism}
Let $(Y, \pi_{XY}, X)$ be the configuration space of a classical field theory. Then, we can define the action of the model as 
\begin{equation}\label{la1}
    S[\phi]:= \int_{X}d^{n}x \, (j^{1}\phi)^{\ast}L,
\end{equation}
with $L: J^{1}Y \rightarrow \mathbb{R}$ being the Lagrangian function. \\
Now, the main object of interest within the geometric covariant formulation is not the Lagrangian function but the so-called Poincare-Cartan forms, $\Theta^{(L)} \in \Omega^{n}(J^{1}Y)$ and  $\Omega^{(L)} \in \Omega^{n+1}(J^{1}Y)$, which satisfy $\Omega^{(L)}:=-d\Theta^{(L)}$. These differential forms on $J^{1}Y$ are given in coordinates by 
\begin{align}\label{pcfl1}
    \Theta^{(L)}&:=\frac{\partial L}{\partial y_{\mu}^{a}} dy^{a}\wedge d^{n-1}x_{\mu} + \left(L -  \frac{\partial L}{\partial y_{\mu}^{a}} y_{\mu}^{a} \right)d^{n}x, \nonumber \\
        \Omega^{(L)}&:=dy^{a}\wedge d\left( \frac{\partial L}{\partial y_{\mu}^{a}} \right) \wedge d^{n-1}x_{\mu} - d\left(L -  \frac{\partial L}{\partial y_{\mu}^{a}} y_{\mu}^{a} \right)\wedge d^{n}x,
\end{align}
where $d^{n-1}x_{\mu}: = \partial_{\mu}\lrcorner d^{n}x$. 
\begin{definicion}
    The pair $(J^{1}Y, \Omega^{(L)})$ is said to be a Lagrangian system. The Lagrangian system and the Lagrangian function are said to be regular if $\Omega^{(L)}$ is 1-nondegenerate. Otherwise, they are singular. 
\end{definicion}
The regularity condition is locally equivalent to 
\begin{equation}\label{rgc1}
    det \left( \frac{\partial^{2} L}{\partial y_{\mu}^{a} \partial y_{\nu}^{b}}\right) \neq 0, \, , \forall \gamma \in J^{1}Y
\end{equation}

Furthermore, the variational problem for a Lagrangian system means the search of the critical sections of the action principle $\eqref{la1}$, which leads to the formulation of Hamilton's principle of the Lagrangian formalism~\cite{romanroy1}. But within the covariant-Lagrangian approach, there is a very useful equivalence that involves the Euler-Lagrange field equations and the Poincare-Cartan $(n+1)$-form, as shown in~\cite{romanroy1}. More precisely, given $(J^{1}E, \Omega^{(\mathcal{L})})$ a Lagrangian system and $\phi \in \mathscr{Y}_{X}$ a critical point for the action principle $\eqref{la1}$ the condition 
\begin{equation}\label{lat1}
(j^{1}\phi)^{\ast}\left( X\lrcorner \Omega^{(\mathcal{L})} \right)=0, \, \forall X \in \mathfrak{X}(J^{1}E),
\end{equation}
is completely equivalent to the Euler-Lagrange field equations of the theory.

We are now interested in study the action of the gauge group of the theory under consideration over the configuration space of the system. To this end, let $\mathcal{G}$ be a Lie group and $\mathfrak{g}$ its Lie algebra. Then, for all $\eta \in \mathcal{G}$, the action of $\eta$ on $\pi_{XY}$ is given by a $\pi_{XY}$-bundle automorphism $(\eta_{Y}, \eta_{X})$, where the maps $\eta_{Y}: Y \rightarrow Y$ and $\eta_{X}: X \rightarrow X$ satisfy $\pi_{XY} \circ \eta_{Y}= \eta_{X} \circ \pi_{XY}$. Locally, these transformations read $\eta(x^{\mu}):= \eta^{\mu}_{X}(x^{\nu})$ and $\eta_{Y}(x^{\mu}, y^{a}):= (  \eta^{\mu}_{X}(x^{\nu}), \eta_{Y}^{a}(x^{\nu}, y^{b}))$. Moreover, given the infinitesimal generator of $\eta$, $\xi_{\eta}\in \mathfrak{g}$, we denote $\xi_{\eta}^{X}:= \xi^{\mu}(x^{\nu}) \partial_{\mu} \in \mathfrak{X}^{1}(X)$ and $\xi_{\eta}^{Y}:=\xi^{\mu}(x^{\nu}) \partial_{\mu} + \xi^{a}(x^{\nu}, y^{b}) \partial_{a} \in \mathfrak{X}^{1}(Y)$, where $\partial_{a}:= \frac{\partial}{\partial y^{a}}$, as the infinitesimal generators of the transformations $\eta_{X}$ and $\eta_{Y}$, respectively. Then, the action of $\eta$ can be lifted to $J^{1}Y$ via the map $\eta_{J_{y}^{1}Y}:J_{y}^{1}Y \rightarrow J_{\eta_{Y}(y)}^{1}Y$, which must satisfy $\eta_{J_{y}^{1}Y}(\gamma):= d\eta_{Y}{}_{y} \circ \gamma \circ (d\eta_{X}{}_{x})^{-1}$, $\forall \gamma \in J_{y}^{1}Y$, and whose infinitesimal generator is the first jet prolongation of the vector field $\xi_{\eta}^{Y} \in \mathfrak{X}^{1}(Y) $, $\xi_{\eta}^{J^{1}Y} \in \mathfrak{X}^{1}(J^{1}Y) $, given in coordinates by
\begin{equation}\label{la2}
\xi_{\eta}^{J^{1}Y}=\xi^{\mu}(x^{\nu}) \partial_{\mu} + \xi^{a}(x^{\nu}, y^{b}) \partial_{a}  + \left( \partial_{\mu}\xi^{a}(x^{\nu}, y^{b}) + \partial_{c}\xi^{a}(x^{\nu}, y^{b})y_{\mu}^{c}- \partial_{\mu}\xi^{\sigma}(x^{\nu})y_{\sigma}^{a} \right)\partial_{a}^{\mu}, 
\end{equation}
with $\partial_{a}^{\mu}:= \frac{\partial }{\partial y_{\mu}^{a}}$ denoting the partial derivative with respect to the fiber coordinate $y_{\mu}^{a}$.\\
Thus, as described in~\cite{deleon2}, the Lie group $\mathcal{G}$ is the symmetry group of the theory if for all $\eta \in \mathcal{G}$, the associated infinitesimal transformation $(\eta_{J^{1}Y}, \xi_{\eta}^{J^{1}Y})$ on $J^{1}Y$ satisfies 
\begin{equation}\label{la5}
\mathcal{L}_{\xi_{\eta}^{J^{1}Y}} \Theta^{(L)} = d\alpha_{\eta}^{(L)},
\end{equation}
where $\mathcal{L}_{\xi_{\eta}^{J^{1}Y}}$ symbolizes the Lie derivative along the vector field $\xi_{\eta}^{J^{1}Y} \in \mathfrak{X}^{1}(J^{1}Y)$, while $\alpha_{\eta}^{(L)} \in \Omega^{n-1}_{0}(J^{1}Y)$ is a $\pi_{XJ^{1}Y}$-horizontal $(n-1,0)$-form on $J^{1}Y$, given in coordinates by 
\begin{equation}\label{la6}
\alpha_{\eta}^{(L)}= \alpha_{\eta}^{\mu}(x^{\nu}, y^{a})d^{n-1}x_{\mu}.
\end{equation}
\begin{definicion}
    The pair $(\xi_{\eta}^{J^{1}Y}, \alpha_{\eta}^{(L)})$ related through condition $\eqref{la5}$ will be referred to as a Noether symmetry.
\end{definicion}
With all of this at hand, given the dual pairing $\langle \cdot, \cdot \rangle: \mathfrak{g}^{\ast} \times \mathfrak{g}  \rightarrow \mathbb{R}$, the action of $\mathcal{G}$ on $J^{1}Y$ induces a Lagrangian covariant momentum map $J^{(L)}: J^{1}Y \rightarrow \mathfrak{g}^{\ast} \otimes \bigwedge^{n-1} T^{\ast}J^{1}Y $ such that $\forall \xi_{\eta} \in \mathfrak{g}$
\begin{equation}\label{la7}
dJ^{(L)}(\xi_{\eta})= \xi_{\eta}^{J^{1}Y}\lrcorner \Omega^{(L)},
\end{equation}
where $J^{(L)}(\xi_{\eta}):= \langle J^{(L)}, \xi_{\eta}  \rangle \in \Omega^{n-1}(J^{1}Y)$ is a $(n-1)$-form on $J^{1}Y$ given  explicitly by 
\begin{equation}\label{la8}
J^{(L)}(\xi_{\eta})= \xi_{\eta}^{J^{1}Y}\lrcorner \, \Theta^{(L)}- \alpha_{\eta}^{(L)}. 
\end{equation}
Moreover, the Lagrangian covariant momentum map is important since it allows to construct conserved currents for the solution of the Euler-Lagrange field equations of the system. That is, given $\xi_{\eta} \in \mathfrak{g}$ and $\phi \in \mathscr{Y}_{X} $ a critical point for the action $\eqref{la1}$, the quantity 
\begin{equation}\label{la9}
\mathcal{J}^{(L)}(\xi_{\eta})=(j^{1}\phi)^{\ast}J^{(L)}(\xi_{\eta}),
\end{equation}
correspond to a conserved current of the theory~\cite{deleon2}. To be precise, the last result corresponds to the formulation of the first Noether theorem within the geometric-covariant Lagrangian approach. \\
Now, we will introduce the notion of localizable symmetries in the context of the geometric-cavariant formulation for classical field theories. As discussed in~\cite{deleon2, molgado1, lee1}, a localizable symmetry for a given classical field theory can be defined as follows
\begin{definicion}
A set $\mathcal{C}_{LS}=\{(\xi^{J^{1}Y}, \alpha^{(\mathcal{L})})\}$, where $\xi^{J^{1}Y} \in \mathfrak{X}(J^{1}Y)$ and $\alpha^{(\mathcal{L})} \in \Omega^{n-1}_{0}(J^{1}Y)$ are given by $\eqref{la2}$ and $\eqref{la5}$, respectively, is said to be a set of localizable symmetries if the following conditions hold: 
    \begin{enumerate}
    \item $\mathcal{C}_{LS}$ is a vector space.
    \item Each pair $(\xi^{J^{1}Y}, \alpha^{(\mathcal{L})}) \in \mathcal{C}_{LS}$ is a Noether symmetry. 
    \item For every pair $(\xi^{J^{1}Y}, \alpha^{(\mathcal{L})}) \in \mathcal{C}_{LS}$ and any open sets $U_{1}, U_{2} \subset X$ with disjoint closures there is a pair $(\tilde{\xi}^{J^{1}Y}, \tilde{\alpha}^{(\mathcal{L})}) \in \mathcal{C}_{LS}$ such that 
    \begin{align}
    \xi^{J^{1}Y}(\beta)&= \tilde{\xi}^{J^{1}Y},\,  \alpha^{(\mathcal{L})}(\beta)= \tilde{\alpha}^{(\mathcal{L})}(\beta), \, \forall\beta\in \pi_{XJ^{1}Y}^{-1}(U_{1}),\nonumber \\
        \tilde{\xi}^{J^{1}Y}(\beta)&=0, \tilde{\alpha}^{(\mathcal{L})}(\beta)=0, \, \forall\beta\in \pi_{XJ^{1}Y}^{-1}(U_{2}).
    \end{align}
\end{enumerate}
\end{definicion}
In summary, the above definition says that one can deform the Noether symmetries to zero in the fibers of $J^{1}Y$ over a neighborhood of the base space $X$.

Therefore, if the pair $(\xi_{\eta}^{J^{1}Y}, \alpha_{\eta}^{(L)})$ constitutes a localizable symmetry, the quantity $\eqref{la9}$ gives rise to trivial Lagrangian Noether charges, that is, given $\Sigma_{t} \subset X$ a Cauchy surface of $X$, $(\xi_{\eta}^{J^{1}Y}, \alpha_{\eta}^{(L)})$ a localizable symmetry and $\phi \in \mathscr{Y}_{X} $ a critical point for the action $\eqref{la1}$, the associated Lagrangian Noether charge vanishes, namely 
\begin{equation}
\mathcal{Q}_{\Sigma_{t}}^{(L)}(\xi):= \int_{\Sigma_{t}} (j^{1}\phi \circ \imath_{t})^{\ast}J^{(L)}(\xi)=0,
\end{equation}
where $\imath_{t}: \Sigma_{t} \rightarrow X$ denotes the inclusion map. In fact, the latter corresponds to the second Noether theorem~\cite{marsdem1, lee1}.
\subsection{Multisymplectic formalism}
Next, we will briefly describe the multisymplectic formalism, which happens to be a finite dimensional, geometric and covariant Hamiltonian-like formulation for classical field theory. To this end, given a covariant configuration space $(Y, \pi_{XY}, X)$ for the system under consideration, we define $(Z^{\star}:= \bigwedge_{1}^{n}T^{\ast}Y, \pi_{YZ^{\star}}, Y)$ as the covariant multimomenta phase space of the theory. Locally, an element $\Xi\in Z_{y}^{\star} $ is written as 
\begin{equation}\label{m1}
\Xi:= p d^{n}x + p_{a}^{\mu} dy^{a}\wedge d^{n-1}x_{\mu}, 
\end{equation}
which allows us to identify $(x^{\mu}, y^{a}, p, p_{a}^{\mu})$ as an adapted coordinate system on $Z^{\star}$ and where the fiber coordinates $(p_{a}^{\mu})$ will be referred to as the polymomenta, which, as we will see below, are related to the Lagrangian function of the system, while the coordinates $p$ will be related to the covariant Hamiltonian of the theory. In addition, the composition map $\pi_{XZ^{\star}}:= \pi_{XY} \circ \pi_{YZ^{\star}}$ gives the bundle $(Z^{\star}, \pi_{XZ^{\star}}, X )$. \\
Moreover, we would like to emphasize that the vector space $Z^{\ast}$ is endowed with a canonical $n$-form $\Theta^{(\mathcal{M})} \in \Omega^{n}(Z^{\ast})$, which is locally given by
\begin{equation}\label{m2}
\Theta^{(\mathcal{M})}:= p d^{n}x + p_{a}^{\mu} dy^{a}\wedge d^{n-1}x_{\mu},
\end{equation}
\begin{observacion}
     Although the differential forms in equations $\eqref{m1}$ and $\eqref{m2}$ have the same local representation, the first corresponds to an element of the fiber of the bundle $\pi_{YZ^{\star}}$ over a point $y \in Y $, while the latter stands for a differential $n$-form on the manifold $Z^{\star}$.
\end{observacion}
The differential form defined by equation $\eqref{m2}$ is known as the multisymplectic potential, since it induces the so-called  multisymplectic $(n+1)$-form through the relation $\Omega^{\mathcal{M}}:= - d\Theta^{(\mathcal{M})} \in \Omega^{n}(Z^{\ast})$, which is given explicitly by 
\begin{equation}\label{m3}
\Omega^{(\mathcal{M})}:= dy^{a}\wedge dp_{a}^{\mu} \wedge d^{n-1}x_{\mu}- dp \wedge d^{n}x.
\end{equation}
\begin{definicion}
    The pair $(Z^{\star}, \Omega^{(\mathcal{M})})$ is called a multisymplectic manifold. 
\end{definicion}

Now, in order to study the action of the gauge symmetry group within the multisymplectic approach, let $(Z^{\star}, \pi_{YZ^{\star}}, Y)$ be the multimomenta phase space associated with the classical field theory $\eqref{la1}$. Then, a $\pi_{XZ^{\star}}$-bundle automorphism $(\Phi_{Z^{\star}}, \Phi_{X})$ is said to be a covariant canonical transformation if it preserves the multisymplectic $(n+1)$-form $\eqref{m3}$. It follows that, if $\Phi_{Z^{\star}}$ is the flow associated with the vector field $\xi_{\Phi}^{Z {\star}}\in \mathfrak{X}(Z^{\star})$ then 
\begin{equation}
    \mathcal{L}_{\xi_{\Phi}^{Z {\star} }}\Omega^{(\mathcal{M})}=0.
\end{equation}
We are now in a position to describe how the fiber-preserving transformations on the covariant configuration space, those generated by the symmetry group, induce covariant canonical transformations on the multimomenta phase space of the system. To this end, let $\mathcal{G}$ be the Lie symmetry group of the system and $\mathfrak{g}$ its Lie algebra. Let $\eta \in \mathcal{G}$, then the $\pi_{XY}$-bundle automorphism $(\eta_{Y}, \eta_{X})$ associated with $\eta$ induces a $\pi_{XZ^{\star}}$-bundle automorphism $(\eta_{Z^{\star}}, \eta_{X})$, where $\eta_{Z^{\star}}: Z^{\star}\rightarrow Z^{\star}$ is the canonical lift of $\eta_{Y}$ to $Z^{\star}$ defined by $\eta_{Z^{\star}}(\Xi):= \eta_{Y}^{-1}{}^{\ast}(\Xi)$, $\forall \Xi\in Z^{\star}$. Thus, given the infinitesimal generators of the transformations $\eta_{X}$ and $\eta_{Y}$, $\xi_{\eta}^{X} \in \mathfrak{X}(X)$ and $\xi_{\eta}^{Y} \in \mathfrak{X}(Y)$, respectively, the vector field $\xi_{\eta}^{Z^{\star}} \in \mathfrak{X}(Z^{\star})$ on $Z^{\star}$ given by 
\begin{align}\label{m4}
    \xi_{\eta}^{Z^{\star}} &:= \xi^{\mu}(x^{\nu}) \partial_{\mu} + \xi^{a}(x^{\nu}, y^{b}) \partial_{a} - \left( p \partial_{\mu}\xi^{\mu}(x^{\nu}) + p_{a}^{\mu} \partial_{\mu} \xi^{a}(x^{\nu}, y^{b}) \right) \partial_{p} \nonumber \\
       &- \left(p_{c}^{\mu} \partial_{a}\xi^{c}(x^{\nu}, y^{b})  - p_{a}^{\sigma} \partial_{\sigma}\xi^{\mu}(x^{\nu}) +  p_{a}^{\mu} \partial_{\sigma}\xi^{\sigma}(x^{\nu}) \right) \partial_{\mu}^{a}
\end{align}
corresponds to the infinitesimal generator of the transformation $\eta_{Z^{\star}}$, where $\partial_{p}:= \frac{\partial }{\partial p}$ and $\partial_{\mu}^{a}:= \frac{\partial }{\partial p_{a}^{\mu}}$.\\
As discussed in~\cite{deleon2}, given a Noether symmetry $(\xi_{\eta}^{J^{1}Y}, \alpha_{\eta}^{(L)})$ associated with some $\eta \in \mathcal{G}$, we define the so-called $\alpha_{\eta}^{(\mathcal{M})}$-lift of $\xi_{\eta}^{J^{1}Y} \in \mathfrak{X}(Y)$ to $Z^{\star}$ as the unique vector field $\xi_{\eta}^{\alpha} \in \mathfrak{X}(Z^{\star})$ that projects by means of $d\pi_{YZ^{\star}}:TZ^{\star} \rightarrow TY$ onto $\xi_{\eta}^{Y}$ and also satisfies the condition 
\begin{equation}\label{m5}
    \mathcal{L}_{\xi_{\eta}^{\alpha} }\Omega^{(\mathcal{M})}=d\alpha_{\eta}^{(\mathcal{M})},
\end{equation}
where $\alpha_{\eta}^{(\mathcal{M})} \in \Omega^{n-1}_{0} (Z^{\star})$, locally given by 
\begin{equation}\label{m6}
    \alpha_{\eta}^{(\mathcal{M})}= \alpha_{\eta}^{\mu}(x^{\nu}, y^{a}) d^{n-1} x_{\mu}. 
\end{equation}
Note that the components of $\alpha_{\eta}^{(L)}$ in $\eqref{la6}$ and $\alpha_{\eta}^{(\mathcal{M})}$ here are the same. \\
Moreover, given $\eta \in \mathcal{G}$ the pair $(\eta_{Z^{\star}}^{\alpha}, \xi_{\eta}^{\alpha})$ represents the action of $\eta$ on $Z^{\ast}$. Thus, since the $\alpha^{(\mathcal{M})}$-lifts give rise to covariant canonical transformations, the action of $\mathcal{G}$ on $Z^{\ast}$ has an associated covariant momentum map $J^{(\mathcal{M})}: Z^{\ast} \rightarrow \mathfrak{g}^{\ast} \otimes \bigwedge^{n-1}T^{\ast}Z^{\star}$ such that for all $\xi_{\eta}\in \mathcal{G}$
\begin{equation}\label{m7}
    J^{(\mathcal{M})}(\xi_{\eta}):= \xi_{\eta}^{\alpha}\lrcorner\Theta^{(\mathcal{M})}- \alpha_{\eta}^{(\mathcal{M})}. 
\end{equation}
As we will see below, the zero level set of this covariant momentum map $\eqref{m7}$ will be associated with the set of first-class constraints of the theory. \\
Finally, given $L:J^{1}Y \rightarrow \mathbb{R}$ the Lagrangian function of the classical field theory $\eqref{la1}$, we have that the affine jet bundle and the covariant multimomenta phase space may be related through the covariant Legendre transformation $\mathbb{F}\mathcal{L}: J^{1}Y \rightarrow Z^{\star}$ given locally as 
\begin{equation}\label{m8}
    \mathbb{F}\mathcal{L}(x^{\mu}, y^{a}, y_{\mu}^{a}):= \left( x^{\mu},\, y^{a},\, p= L - \frac{\partial L}{\partial y_{\mu}^{a}} y_{\mu}^{a},\, p_{a}^{\mu}=\frac{\partial L}{\partial y_{\mu}^{a}}  \right). 
\end{equation}
Then, by using the covariant Legendre transformation $\eqref{m8}$, we may induce information of the classical field theory under consideration from the affine jet bundle to the covariant multimomenta phase space and vice versa. In particular, it is not difficult to see that, using the Legendre transformation $\eqref{m8}$, we can obtain the the Poincare-Cartan forms $\eqref{pcfl1}$ by means of the relations~\cite{deleon1}
\begin{align}
    \Theta^{(L)}&= \mathbb{F}\mathcal{L}^{\ast}\Theta^{(\mathcal{M})}, \nonumber\\
    \Omega^{(L)}&= \mathbb{F}\mathcal{L}^{\ast}\Omega^{(\mathcal{M})}.
\end{align}
\subsection{Polysymplectic analysis}\label{ssp}
In this subsection we will introduce the polysymplectic approach for classical field theory, which is a covariant Poisson-Hamiltonian framework. To this end, note that the multimomenta phase space $(Z^{\star}:= \bigwedge_{1}^{n}T^{\ast}Y, \pi_{YZ^{\star}}, Y)$ has as a subbundle the vector bundle $(\bigwedge{}_{0}^{n}T^{\ast}Y, \pi_{Y\bigwedge_{0}^{n}T^{\ast}Y}, Y)$, which allows us to introduce a new vector bundle over $Y$, given by the quotient bundle 
\begin{equation}\label{ps1}
    P:= Z^{\ast}/\bigwedge{}_{0}^{n}T^{\ast}Y.
\end{equation}
An element of the fiber $\vartheta \in P_{y}$ at $y$ can be written locally as 
\begin{equation}\label{ps2}
    \vartheta=p_{a}^{\mu}dy^{a}\wedge d^{n-1}x_{\mu}, 
\end{equation}
which allows us to identify $(x^{\mu}, y^{a}, p_{a}^{\mu})$ as an adapted coordinate system on $P$. Note that, the composition map $\pi_{XP}=\pi_{XY} \circ \pi_{YP}$ gives rise to the polymomenta phase-space $(P, \pi_{XP}, X)$. Let $\mathscr{P}_{X}$ be the set of sections of $\pi_{XP}$, then, a section $\rho \in \mathscr{P}_{X}$ such that $\phi= \pi_{YP} \circ \rho \in \mathscr{Y}_{X}$ can be locally represented by $(x^{\mu}, \phi(x^{\mu}), \pi_{a}^{\mu}(x^{\mu}))$. Furthermore, the definition of the quotient space $\eqref{ps1}$ yields the bundle structure $(Z^{\star}, \pi_{PZ^{\star}}, P)$, where a section $h \in \mathscr{Z}^{\star}_{P}$ is called Hamiltonian if it locally reads 
\begin{equation}
    h(x^{\mu}, y^{a}, p_{a}^{\mu})=(x^{\mu}, y^{a}, p=-H(x^{\mu}, y^{a}, p_{a}^{\mu}), p_{a}^{\mu}),
\end{equation}
being $H$ the Hamiltonian function associated with $h$.\\
There is a relation between $J^{1}Y$ and $P$ given by the covariant Legendre map $\mathbb{F}\mathcal{L}_{DW}: J^{1}Y \rightarrow P$ defined as 
\begin{equation}\label{ps3}
    \mathbb{F}\mathcal{L}_{DW}(x^{\mu}, y^{a}, p_{a}^{\mu}):=\left( x^{\mu}, y^{a}, p_{a}^{\mu}= \frac{\partial L}{\partial y_{\mu}^{a}}  \right), 
\end{equation}
which allows us to define the De Donder-Weyl Hamiltonian section $h_{DW} \in  \mathscr{Z}^{\star}_{P}$, that is, a section satisfying $h_{DW} \circ  \mathbb{F}_{DW}\mathcal{L} =  \mathbb{F}\mathcal{L}$ and whose Hamiltonian function is identified with the De Donder-Weyl Hamiltonian 
\begin{equation}\label{ps4}
    H_{DW}(x^{\mu}, y^{a}, p_{a}^{\mu}):= p_{a}^{\mu} y_{\mu}^{a} - L(x^{\mu}, y^{a}, p_{a}^{\mu}).
\end{equation}
Now, in the polymomenta phase-space $\pi_{XP}$ we can define the so-called polysymplectic structure, as an equivalence class of forms $\Omega^{V}$ given by 
\begin{equation}\label{ps5}
    \Omega^{V}:= [dp_{a}^{\mu}\wedge dy^{a} \wedge \bar{\omega}_{\mu} \, \text{mod} \, \Omega^{n+1}_{1}(P)],
\end{equation}
where $\bar{\omega}_{\mu}=\partial_{\mu}\lrcorner\bar{\omega}$, being $\bar{\omega}:=dx^{0} \wedge \dotsc \wedge dx^{n-1}$ the horizontal volume form on $P$. Let $X \in \mathfrak{X}^{p}(P)$ be a vertical $p$-multivector field, then we say a horizontal $(n-p; 0)$-form $F \in \Omega^{n-p}_{0}(P)$ is a Hamiltonian form if it satisfies the condition 
\begin{equation}\label{ps6}
    X\lrcorner\Omega^{V}= d^{V}F,
\end{equation}
where $d^{V}: \Omega^{p}_{q} (P) \rightarrow \Omega^{p+1}_{q+1}(P)$ stands for the vertical exterior derivative. We will denote the set of Hamiltonian $p$-forms on $P$ as $\Omega^{p}_{H}(P)$. It is important to mention that the space $\Omega^{p}_{H}(P)$ is not stable with respect to the exterior product, then the natural product operation of Hamiltonian forms is the \textit{co-exterior} product
\begin{equation}\label{ps7}
	F \bullet G:= \ast^{-1}(\ast F \wedge \ast G)
\end{equation}
being $\ast: \Omega^{p}(X) \rightarrow\Omega^{n-p}(X)$ is the dual Hodge operator on the base manifold $X$. As a consequence, the product $\bullet: \Omega^{p}_{H}(P) \times \Omega^{q}_{H}(P) \rightarrow \Omega^{p+q-n}_{H}(P)$ is graded commutative and associative.\\
The Poisson-Gerstenhaber bracket $\{[\cdot, \cdot ]\}: \Omega^{p}_{H}(P) \times \Omega^{q}_{H}(P) \rightarrow \Omega^{p+q+1-n}_{H}(P)$ is defined by 
\begin{equation}\label{ps8}
    \{[ F, G ]\}:= (-1)^{n-p} X_{F}\lrcorner X_{G}\lrcorner \Omega^{V},
\end{equation}
where $X_{F}$ and $X_{G}$ are multivector fields associated with $F \in \Omega^{p}_{H}(P)$ and $G \in \Omega^{q}_{H}(P)$, respectively. In fact, the Poisson-Gerstenhaber bracket $\eqref{ps8}$ is induced by the Schouten-Nijenhuis bracket of the corresponding Hamiltonian multivector fields~\cite{kana3}. Furthermore, as demonstrated in~\cite{kana4, kana3}, the space $ \Omega^{p}_{H}(P)$ of Hamiltonian forms with the operations $\{[\cdot, \cdot ]\}: \Omega^{p}_{H}(P) \times \Omega^{q}_{H}(P) \rightarrow \Omega^{p+q+1-n}_{H}(P)$ and $\bullet: \Omega^{p}_{H}(P) \times \Omega^{q}_{H}(P) \rightarrow \Omega^{p+q-n}_{H}(P)$ is a Poisson-Gerstenhaber algebra, that is,
\begin{align}\label{ps9}
    \{[F, G]\}&=-(-1)^{g_{1}g_{2}} \{[G, F]\} \nonumber\\
			\{[F, G \bullet H]\}&= \{[F, G ]\} \bullet H + (-1)^{g_{1}(g_{2} + 1)} F\bullet \{[G, H]\}\nonumber \\
			(-1)^{g_{1}g_{3}} \{[F, \{[G, H ]\}]\} &+ (-1)^{g_{1}g_{2}} \{[G, \{[H, F ]\}]\} + (-1)^{g_{2}g_{3}} \{[H, \{F, G ]\}]\} =0,
\end{align}
where $g_{1}=n-p-1$, $g_{2}=n-q-1$, $g_{3}=n-p-1$.\\
Note that the Poisson-Gerstenhaber bracket enable us to identify the pairs of canonically conjugate variables
\begin{align}\label{ps10}
    \{[ p_{a}^{\mu} \omega_{\mu}, y^{b}\omega_{\nu}]\}&=\delta_{a}^{b} \omega_{\nu}.
\end{align}
In consequence, given a section $\varrho \in \mathscr{P}_{X}$ we can write the so-called De Donder-Weyl-Hamiltonian equations in terms of the Poisson-Gerstenhaber bracket as 
\begin{equation}\label{ps11}
    \begin{split}
        \partial_{\mu}\phi^{a}&= \varrho^{\ast} \{[ H_{DW}, y^{a}\bar{\omega}_{\mu} ]\}\\
         \partial_{\mu}\pi_{a}^{\mu}&= \varrho^{\ast} \{[ H_{DW}, p_{a}^{\mu}\bar{\omega}_{\mu} ]\},
    \end{split}
\end{equation}
which are completely equivalent to the Euler-Lagrange field equations, as long as the theory is non-singular. In addition, we can write the field equations associated with an arbitrary Hamiltonian $(n-1)$-form in terms of the Poisson-Gerstenhaber bracket. In order to do so, let $d\bullet: \Omega_{H}^{p} (P)\rightarrow \Omega_{H}^{(n-1)-p}(X)$ be the total co-exterior derivative and $F \in \Omega_{H}^{p}$ then the field equations take the form 
\begin{equation}\label{ps12}
	d\bullet F = -\sigma(-1)^{n} \varrho^{\ast} \{[H_{DW}, F]\} + d^{h}\bullet F, 
\end{equation}
where $\sigma= \pm 1$ stand for the signature in the base space manifold $X$, and the total co-exterior derivative $d\bullet: \Omega_{H}^{p} (P)\rightarrow \Omega_{H}^{(n-1)-p}(X)$ and the horizontal co-exterior derivative $d^{h}\bullet: \Omega_{H}^{p} (P)\rightarrow \Omega_{H}^{(n-1)-p}(X)$ are explicitly given by 
\begin{align}\label{ps13}
    d\bullet F &= \frac{1}{(n-p)!}\left( \varrho^{\ast}d\left( F \bullet dx^{\mu_{1}}\wedge \cdots \wedge dx^{\mu_{n-p}}\right) \right) \bullet \bar{\omega}_{\mu_{1}\ldots \mu_{n-p}},\nonumber \\
	d^{h}\bullet F &= \frac{1}{(n-p)!}\left( \varrho^{\ast} \partial_{\sigma}\left( F \bullet dx^{\mu_{1}}\wedge \cdots \wedge dx^{\mu_{n-p}}\right) dx^{\sigma}\right) \bullet \bar{\omega}_{\mu_{1}\ldots \mu_{n-p}},
\end{align}
being $ \bar{\omega}_{\mu_{1}\ldots \mu_{n-p}}= \partial_{\mu_{1}}\lrcorner \cdots \partial_{\mu_{n-p}}\lrcorner \bar{\omega}$ the basis for the horizontal $(p; 0)$-forms on $P$. \\
Now, we briefly describe the treatment for singular system within the polysymplectic approach, as discussed in~\cite{kana4}. To this end, we say that a system is singular if it does not satisfy the regularity condition $\eqref{rgc1}$. In this case, the covariant Legendre map $\eqref{ps3}$ is not invertible, then it can not be a diffeomorphism and in consequence there are conditions emerging from the definition of the polymomenta $p_{a}^{\mu}= \frac{\partial L}{\partial y_{a}^{\mu}}$, namely,
\begin{equation}\label{ps14}
	C^{(k)\mu}(y^{a}, p_{a}^{\nu}) \approx 0
\end{equation}
which, as in Dirac's terminology, will be called primary constraints to emphasize that the field equations are not used to obtain these relations and that they imply no restriction on the field variables and its derivatives. Here, the symbol $\approx$ is used to specify that a certain relation is valid only on the surface of the polymomenta phase-space defined by the constraints of the system, that is, on the constraint surface. Moreover, in this present approach we restrict our attention to the constraints which can be recognized as Hamiltonian $(n-1)$-forms 
\begin{equation}\label{ps15}
	C^{(k)}:= C^{(k)\mu} \bar{\omega}_{\mu},
\end{equation}
where the index $k$ runs over the complete set of primary constraints $(n-1)$-forms, and  $\bar{\omega}_{\mu}$ stands for the horizontal volume form on $P$, as described below $\eqref{ps5}$.\\
Now, in analogy with the usual Dirac's Hamiltonian algorithm for singular Lagrangian systems~\cite{henb1}, we need to introduce the total De Donder-Hamiltonian function $\bar{H}_{DW}: P \rightarrow \mathbb{R}$, which is defined by 
\begin{equation}\label{ps16}
	\bar{H}_{DW}=H_{DW} + u_{(k)}\bullet C^{(k)},
\end{equation}
where the Lagrange multipliers $u_{(k)} \in \Omega_{0}^{1}(P)$. Note that, $ \bar{H}_{DW}$ and $H_{DW}$ are equivalent on the constraint surface, but in order to obtain the correct field equations of the theory under consideration, we will implement the total Hamiltonian $\bar{H}_{DW}$ within our analysis. \\
In addition, as described in~\cite{kana4}, the necessary condition to analyse singular Lagrangian is the conservation of each the constraint $(n-1)$-form under the co-exterior derivative, namely 
\begin{equation}\label{ps17}
	d\bullet C^{(k)}= \{[H_{DW}, C^{(k)} ]\} + u_{(l)} \bullet \{[C^{(l)}, C^{(k)}  ]\} \approx  0,
\end{equation}
which is analogous to the consistency conditions within the Dirac-Hamiltonian approach. Note that equation $\eqref{ps17}$ can either impose a restriction on the Lagrange multipliers $u$'s or it may reduce to a new relation independent of the $u$'s and the primary constraints. In the latter case, following the Dirac's terminology, these new relations will be referred to as secondary constraints. Thus, if there are secondary constrains, after writing it as $(n-1)$-forms $B^{(l)}:=B^{(l)\mu} \bar{\omega}_{\mu}$ (where $l$ runs over the complete set of secondary constraint $(n-1)$-forms), we must again impose a consistency condition of the form $\eqref{ps17}$, which must imply new secondary constraints or whether it only fixes the Lagrange multipliers $u_{(l)}$ and so on. It is similar to the usual procedure within the Dirac formalism. Hence, after finishing the process of generating new constraints, we will have the complete set of constraints $(n-1)$-forms $\{ C^{(m)} \}$ (here the index $m$ runs over the complete set of primary, secondary, etc., constraints $(n-1)$-forms) characterizing the singular system under study.\\
Now, in analogy with the Dirac approach, we may introduce the notion of first- and second-class Hamiltonian forms within the polysymplectic formalism. 
\begin{definicion}
	A Hamiltonian $p$-form $F \in \Omega_{H}^{p}(P)$ is said to be first-class if its Poisson-Gerstenhaber bracket with every constraints $(n-1)$-form $C^{(m)}$ weakly vanishes, 
\begin{equation}\label{ps18}
		\{[F, C^{(m)}]\}\approx 0.
\end{equation}
Otherwise, they are said to be second-class.
\end{definicion}
Thus, we can classify the complete set of constraints $(n-1)$-forms $C^{(m)}$ into subsets of first- and second-class constraints, denoted by $\{\mathcal{C}^{(i)}\}$ and $\{\mathcal{B}^{(i)}\}$, respectively (where the index runs over the set of fist- and second-class constraints $(n-1)$-forms). \\
Second-class constraints are presented whenever the $(n-1)$-form valued matrix 
\begin{equation}\label{ps19}
	\mathcal{B}^{(i,j)}:= \{[\mathcal{B}^{(i)}, \mathcal{B}^{(j)}]\}
\end{equation}
does not vanish on the constraint surface. For simplicity, we assume that the constraints are irreducible and the rank of $\mathcal{B}^{(i,j)}$ is constant on the constraints surface. Moreover, since $\mathcal{B}^{(i,j)}$ is a nonsingular matrix whose components are $(n-1)$-forms, its $(1;0)$-form valued inverse matrix $\mathcal{B}_{(i,j)}^{-1}$ exists such that 
\begin{equation}\label{ps20}
	\mathcal{B}^{-1}{}_{(i,j)} \wedge \mathcal{B}^{(j,k)} = \delta_{i}^{k} \bar{\omega}. 
\end{equation}
Bearing this in mind and in analogy with the Dirac formalism for singular systems, we can define a Dirac-Poisson bracket for Hamiltonian $0$- or $(n-1)$-forms on the polymomenta phase space. 
\begin{definicion}
Let $F$ be a Hamiltonian $0$- or $(n-1)$-form and $G$ a Hamiltonian $(n-1)$-form, then a Dirac-Poisson bracket within the polysymplectic approach can be defined as 
\begin{equation}\label{ps21}
	\{[F, G]\}_{D}:= \{[F, G]\}- \sigma \{[F, \mathcal{B}^{(i)}]\} \bullet \left( \mathcal{B}^{-1}{}_{(i,j)}\wedge \{[\mathcal{B}^{(j)}, G]\} \right). 
\end{equation}
\end{definicion}
As demonstrated in~\cite{kana4}, the Dirac-Poisson bracket of any Hamiltonian $(n-1)$-form with a second class constraint vanishes. From our point of view, this fact is important, since we can use the Dirac-Poisson bracket, instead of the Poisson-Gerstenhaber bracket, to obtain the correct field equations while taking the second-class constraint $(n-1)$-forms of the system as strong identities. Furthermore, the Dirac-Poisson bracket satisfies the expected properties of a Lie algebra. Finally, we suggest the reader to reference~\cite{kana4} for further details on the construction of the bracket $\eqref{ps21}$ and its properties.\\

In the following subsection, we will describe how to recover the instantaneous Dirac-Hamiltonian formulation for a given classical field theory from the point of view of the multisymplectic formalism.

\subsection{Space plus time decomposition for classical field theory}
In this section we briefly describe the space plus time decomposition of the geometric-covariant Lagrangian and multisymplectic formulations for classical field theory, closely following~\cite{gotay1, gotay2, molgado1, deleon1, molgado3}. To this end, let $\Sigma$ be a compact (oriented, connected) $(n-1)$-manifold without boundary and we define a slicing of the spacetime manifold $X$ as a diffeomorphism $\Psi: \mathbb{R} \times \Sigma \rightarrow X$. We denote by $\mathscr{X}= \{\Psi_{t} : \Sigma \rightarrow X| t\in \mathbb{R}\}$ the space of all smooth embeddings of $\Sigma$ into $X$, which define the Cauchy surfaces by $\Sigma_{t}= \Psi_{t}(\Sigma)$. Let $\partial_{t} \in \mathfrak{X}(\mathbb{R} \times \Sigma)$ be the infinitesimal generator of the translations on $\mathbb{R} \times \Sigma$. Then, we define $\zeta^{X}:=d\Psi (\partial_{t})$ as the infinitesimal generator of $\Psi: \mathbb{R} \times \Sigma \rightarrow X$ which is, by definition, everywhere transverse to $\Sigma_{t}$. Considering Cauchy surfaces locally given by a level set of the coordinate $x^{0}$, a coordinate system on $\Sigma_{t}$ can be simply denoted by $(x^{i})$, $i \in \{1,\ldots, n-1 \}$. \\
Now, given a fibre bundle $\pi_{XK}: K \rightarrow X$ and a slicing $\Psi$ of $X$, a compatible slicing of $K$ is a bundle $\pi_{\Sigma K_{\Sigma}}: K_{\Sigma} \rightarrow \Sigma$ together with a bundle diffeomorphism $\Psi_{K}:\mathbb{R}\times K_{\Sigma} \rightarrow K$ such that the diagram
\begin{center}
	$\xymatrix{ \mathbb{R}\times K_{\Sigma}  \ar[d] \ar[r]^{\Psi_{K}} & K \ar[d] \\ \mathbb{R}\times \Sigma \ar[r]^{\Psi} & X  }$
\end{center}
commutes, where the vertical arrows are bundle projections. Note that, for $t\in \mathbb{R}$, $\Psi_{K,t}: K_{\Sigma} \rightarrow K$ defines an embedding such that $K_{t}:=\Psi_{K,t}(K_{\Sigma}) \subset K$. Moreover, given $\partial_{t} \in \mathfrak{X}(\mathbb{R} \times K_{\Sigma})$, the infinitesimal generator of $\Psi_{K}:\mathbb{R}\times K_{\Sigma} \rightarrow K$ is given by $\zeta^{K}:=d\Psi_{K}(\partial_{t})$ which projects to $\zeta^{X}$ by means of $d\pi_{XK}: TK \rightarrow TX$ and is also everywhere transverse to $K_{t}$.  Bearing this in mind, we can introduce the following 
\begin{definicion}
	The pairs $(\Sigma_{t}, \zeta^{X})$ and $(K_{t}, \zeta^{K})$ are said to be an infinitesimal and compatible slicing of $\pi_{XK}$.\footnote{These pairs, in turn define an one-parametric group of $\pi_{XK}$-bundle automorphisms}
\end{definicion}
In addition, the restriction of the bundle $\pi_{XK}$ to $\Sigma_{t}$, that is, $\pi_{\Sigma_{t} K_{t}} : K_{t} \rightarrow \Sigma_{t}$, has as an adapted coordinate system $(x^{i}, k^{a})$, where $k^{a}$ are fibre coordinates on $K$ with $a \in \{1, \ldots, m^{\prime}\}$.\\

Now, let $\mathscr{K}_{t}:= \{\phi: \Sigma \rightarrow K_{t} | \pi_{\Sigma_{t} K_{t}} \circ \phi = id_{\Sigma_{t}}\}$ be the set of sections of the fiber bundle $\pi_{\Sigma_{t} K_{t}}$, then the collection
\begin{equation}\label{bspt1}
	\mathscr{K}^{\Sigma}:= \bigcup_{\Psi_{t} \in \mathscr{X}} \mathscr{K}_{t},
\end{equation}
defines an infinite-dimensional fiber bundle $(\mathscr{K}^{\Sigma}, \pi_{\mathscr{X} \mathscr{K}^{\Sigma}},  \mathscr{X})$ over $\mathscr{X}$. At $\kappa \in \mathscr{K}_{t}$ the induced coordinates on $\mathscr{K}_{t}$ are $\kappa^{a}=k^{a} \circ \kappa$, being $(k^{a})$ coordinates along the fibers of $K \rightarrow X$. 
\begin{definicion}
The tangent space to $\mathscr{K}_{t}$ at $\kappa \in \mathscr{K}_{t}$ is 
\begin{equation}\label{bspt2}
	T_{\kappa}\mathscr{K}_{t}:= \{ \dot{\kappa}: \Sigma_{t}  \rightarrow V K_{t} | \pi_{K_{t}VK_{t}} \circ \dot{\kappa}= \kappa\},
\end{equation}
where $\pi_{K_{t}VK_{t}}: VK_{t} \rightarrow K_{t}$ is the restriction of the vertical tangent bundle $\pi_{KVK}: VK \rightarrow K$ to $K_{t}$.
\end{definicion}
Locally, an element $\dot{\kappa} \in T_{\kappa}\mathscr{K}_{t}$ can be written as 
\begin{equation}\label{bspt3}
	\dot{\kappa}= \dot{\kappa}^{a} \frac{\partial }{\partial \kappa^{a}},
\end{equation}
where the functions $\dot{\kappa}^{a}$ depend on the coordinates on the Cauchy surface $\Sigma_{t}$. \\
Similarly, we can define the cotangent space to $\mathscr{K}_{t}$ by 
\begin{definicion}
	The cotangent space to $\mathscr{K}_{t}$ at $\kappa \in \mathscr{K}_{t}$ is defined by
	\begin{equation}\label{bspt4}
		T^{\ast}_{\kappa}\mathscr{K}_{t}:=\{ \tau: \Sigma_{t} \rightarrow L(VK_{t}, \bigwedge^{n-1}\Sigma_{t}): \pi_{\mathscr{K}_{t} L(VK_{t}, \bigwedge^{n-1}\Sigma_{t})} \circ \tau = \kappa\},
	\end{equation}
where $ L(VK_{t}, \bigwedge^{n-1}\Sigma_{t})$ is the vector bundle over $K_{t}$ whose fibre above $k \in \mathscr{K}_{t}$ is the set of linear maps from $VK_{t}$ to $ \bigwedge^{n-1}\Sigma_{t}$.
\end{definicion}
Note that, in adapted coordinates, an element $\tau \in T^{\ast}_{\kappa}\mathscr{K}_{t}$ can be expressed as
\begin{equation}\label{bspt5}
    \tau= \tau_{a} d\kappa^{a}\otimes d^{n-1}x_{0},
\end{equation}
where $d^{n-1}x_{0}$ denotes the $(n-1)$-form on the Cauchy surface $\Sigma_{t}$. Hence, the natural pairing between elements $\dot{\kappa} \in T_{\kappa}\mathscr{K}_{t}$ and $\tau \in T^{\ast}_{\kappa}\mathscr{K}_{t}$ is given by integration, specifically
\begin{equation}\label{bspt6}
    \langle \dot{\kappa}, \tau \rangle:= \int_{\Sigma_{t}} \dot{\kappa} \lrcorner \tau. 
\end{equation}

Finally, given $\alpha \in \Omega^{q+n-1}(K)$ a differential form on $K$, we can define an associated differential $q$-form $\alpha_{t} \in \Omega^{q}(\mathscr{K}_{t})$ on $\mathscr{K}_{t}$
\begin{definicion}
    Given $\alpha \in \Omega^{q+n-1}(K)$ there is $\alpha_{t} \in \Omega^{q}(\mathscr{K}_{t})$ such that 
    \begin{equation}\label{bspt7}
        \alpha_{t} (\sigma)(V_{1},\ldots, V_{q}):= \int_{\Sigma_{t}}\sigma^{\ast}(V_{q}\lrcorner\cdots V_{1}\lrcorner\alpha),
    \end{equation}
    where $\sigma \in \mathscr{K}_{t}$ and $V_{1}, \ldots,V_{q}\in T_{\sigma}\mathscr{K}_{t}$.
\end{definicion}
As we will see below, the above definition of a differential form on $\mathscr{K}_{t}$ will be useful for inducing a presymplectic form on the set of sections of a certain restriction of the covariant multimomenta phase space $\pi_{YZ^{\star}}$.\\

With the above space plus time decomposition of fiber-bundles at hand, we are in position to relate the finite- and infinite-dimensional formulations for a given classical theory. To this end, as discussed in~\cite{gotay1, molgado1}, let $(\Sigma_{t}, \zeta^{X})$ and $(Y_{t}, \zeta^{Y})$ be a $\mathcal{G}$-slicing of the covariant configuration space $\pi_{XY}$.\footnote{Namely an infinitesimal and compatible slicing of $\pi_{XY}$ such that for $\xi_{\eta} \in \mathfrak{g}$ the first jet prolongation $\zeta^{J^{1}Y}=j^{1}\zeta^{Y} \in \mathscr{X}(J^{1}Y)$ of $\zeta^{Y} \in \mathscr{X}(Y)$ defines an infinitesimal symmetry of the theory} Let $\pi_{\Sigma_{t}Y_{t}}$ be the  restriction of the bundle $\pi_{XY}$ to $\Sigma_{t}$ whose adapted coordinate system on $Y_{t}$ can be identified as $(x^{i}, y^{a})$. Furthermore, let $\mathscr{Y}_{t}$ be the set of sections of $\pi_{\Sigma_{t}Y_{t}}$ and note that for all $\varphi \in \mathscr{Y}_{t}$ there is a section $\phi\in \mathscr{Y}_{X}$ of $\pi_{XY}$ such that $\varphi = \phi \circ \imath_{t}$, where $\imath_{t}: \Sigma_{t} \rightarrow X$ denotes the inclusion map. Now, introducing $\mathscr{Y}_{t}$ as the $t$-instantaneous configuration space of the classical field theory $\eqref{la1}$, we can identify $T\mathscr{Y}_{t}$ as the $t$-instantaneous space of velocities whose adapted coordinate system is denoted by $(\varphi^{a}, \dot{\varphi}^{a})$, where the temporal derivative coordinate is given by 
\begin{equation}\label{bspt8}
    \dot{\varphi}^{a}:= \left. (\mathcal{L}_{\zeta^{Y}} \phi)^{q} | \right._{\Sigma_{t}}= \left. (d\phi \circ \zeta^{X} - \zeta^{Y} \circ \phi)^{a} |\right._{\Sigma_{t}}
\end{equation}
with $\phi \in \mathscr{Y}_{X}$ and $d\phi: TX \rightarrow TY$. \\
Before proceeding with the construction of the $t$-instantaneous Lagrangian of the system, we need to introduce some notation 
\begin{notacion}
    The triple $((J^{1}Y)_{t}, \pi_{Y_{t}(J^{1}Y)_{t}}, Y_{t})$ denotes the restriction of the affine bundle $\pi_{YJ^{1}Y}$ to $Y_{t}$, meanwhile $(J^{1}Y_{t}, \pi_{Y_{t}J^{1}Y_{t}}, Y_{t})$ denotes the affine jet bundle of $\pi_{\Sigma_{t}Y_{t}}$.
\end{notacion}
Hence, as discussed in~\cite{gotay1}, given $(x^{i}, y^{a}, y_{\mu}^{a})$ an adapted coordinate system on $(J^{1}Y)_{t}$, the jet decomposition map $\beta_{\zeta^{Y}}: (J^{1}Y)_{t} \rightarrow J^{1}Y_{t} \times VY_{t}$ over $Y_{t}$, locally given by $\beta_{\zeta^{Y}}(x^{i}, y^{a}, y_{\mu}^{a}):= (x^{i}, y^{a}, y_{i}^{a},\dot{y}^{a})$, induces an isomorphism between $(\mathscr{J}^{1}\mathscr{Y})_{t}$ and $T \mathscr{Y}_{t}$. Observe that it can be extended to a map on sections; in fact, given $j^{1}\phi$ the first jet prolongation of $\phi \in \mathscr{Y}_{X}$, we can write $\beta_{\zeta^{Y}}(j^{1}\phi \circ \imath_{t}):= (j^{1}\varphi, \dot{\varphi})$, where $j^{1}\varphi$ is the first jet prolongation of $\varphi:= \phi \circ \imath_{t} \in \mathscr{Y}_{t}$. Bearing this in mind, we can obtain the instantaneous Lagrangian density of the theory, $\mathcal{L}_{t,\zeta^{Y}}: J^{1}Y_{t} \times VY_{t} \rightarrow \bigwedge^{n-1}T^{\ast}\Sigma_{t}$ by means of $\mathcal{L}_{t,\zeta^{Y}} (j^{1}\varphi, \dot{\varphi}):= (j^{1}\phi \circ \imath_{t})^{\ast}(\zeta^{X}\lrcorner \mathcal{L})$ and, consequently, the $t$-instantaneous Lagrangian functional $L_{t, \zeta^{Y}}: T\mathscr{Y}_{t} \rightarrow \mathbb{R}$ is given by 
\begin{equation}\label{bspt9}
    L_{t, \zeta^{Y}}(\varphi, \dot{\varphi})=\int_{\Sigma_{t}}L(j^{1}\varphi, \dot{\varphi})\zeta^{0} d^{n-1}x_{0}
\end{equation}
where $\zeta^{0}$ corresponds to the component of the generator $\zeta^{X}$ along $\partial_{0}$ and $d^{n-1}x_{0}$ denotes the $(n-1)$-form on the Cauchy surface $\Sigma_{t}$. It follows that the instantaneous Legendre transformation is defined as a bundle map $\mathbb{F}L_{t, \zeta^{Y}}: T\mathscr{Y}_{t}\rightarrow T^{\ast}\mathscr{Y}_{t}$ over $\mathscr{Y}_{t}$ 
\begin{equation}\label{bspt10}
	\mathbb{F}L_{t, \zeta^{Y}}(\varphi^{a}, \dot{\varphi}^{a}):= \left( \varphi^{a}, \pi_{a} := \frac{\partial L}{\partial  \dot{\varphi}^{a}} \right),
\end{equation}
which, in turn, allows us to identify the $t$-primary constraint set of the theory $\mathcal{P}_{t, \zeta^{Y}} \subset T^{\ast}\mathscr{Y}_{t}$ as the submanifold on the $t$-instantaneous phase space of the system, $T^{\ast}\mathscr{Y}_{t}$, characterized by the image of the instantaneous Legendre transformation $\eqref{bspt10}$. As usual, in adapted coordinates, the symplectic structure $\omega_{t}\in \Omega^{2}(T^{\ast}\mathscr{Y}_{t})$ on $T^{\ast}\mathscr{Y}_{t}$ is given by 
\begin{equation}\label{bspt11}
	\omega_{t}(\varphi, \pi):= \int_{\Sigma_{t}}d\varphi^{a}\wedge d\pi_{a} \otimes d^{n-1}x_{0}.
\end{equation}
Now, we proceed to decompose the multimomenta phase space and introduce a reduction process to relate the multisymplectic and the instantaneous Hamiltonian approaches. To this end, let $(Z_{t}^{\star}, \pi_{Y_{t}Z_{t}^{\star}}, Y_{t})$ be the restriction of $\pi_{YZ^{\star}}$ to $Y_{t}$. 
\begin{notacion}
    $\mathscr{Z}_{t}^{\star}$ denotes the set of sections of the bundle $\pi_{\Sigma_{t}Z_{t}^{\star}}:= \pi_{\Sigma_{t}Y_{t}} \circ \pi_{Y_{t}Z_{t}^{\star}}$
\end{notacion}
In particular, the relation $\eqref{bspt7}$ allows us to endow the space $\mathscr{Z}_{t}^{\star}$ with a presymplectic form $\Omega_{t}$, hence $(\mathscr{Z}_{t}^{\star}, \Omega_{t})$ is a presymplectic manifold. Thus, in order to see that the quotient space $\mathscr{Z}_{t}^{\star} / \text{ker}\,\Omega_{t}$ is canonically isomorphic to $T^{\ast}\mathscr{Y}_{t}$, as explained in~\cite{gotay1}, we need to introduce the bundle map $R_{t}: \mathscr{Z}_{t}^{\star} \rightarrow T^{\ast}\mathscr{Y}_{t}$ over $\mathscr{Y}_{t}$ defined by 
\begin{equation}\label{bspt12}
    \langle R_{t}(\sigma), V \rangle:= \int_{\Sigma_{t}}\varphi^{\ast}(V\lrcorner \sigma),
\end{equation}
where $\varphi:= \pi_{YZ^{\star}} \circ \sigma \in \mathscr{Y}_{t}$ and $V \in T_{\varphi}\mathscr{Y}_{t}$. Locally, it reads $R_{t}(\sigma)= (p_{a}^{0} \circ \sigma)dy^{a} \otimes d^{n-1}x_{0}$, then $\text{ker}\, R_{t}:= \{\sigma \in  \mathscr{Z}_{t}^{\star} | p_{a}^{0} \circ \sigma=0 \}$. Observe that since the symplectic form $\omega_{t}\in \Omega^{2}(T^{\ast}\mathscr{Y}_{t})$ is non-degenerate, then $R_{t}^{\ast}\omega_{t}=\Omega_{t}$, and hence $\text{ker}\, dR_{t}= \text{ker}\, \Omega_{t}$. Consequently, the map $\mathscr{Z}_{t}^{\star} / \text{ker}\,\Omega_{t} \rightarrow T^{\ast}\mathscr{Y}_{t}$ is a symplectic diffeomorphism, as proved in~\cite{gotay1}. Further, the subset $\mathscr{N}_{t}:= \{\sigma \in \mathscr{Z}_{t}^{\star} \left.|\right. \sigma=\mathbb{F}\mathcal{L} \circ j^{1}\phi \circ \imath_{t}\} \subset \mathscr{Z}_{t}^{\star} $ projects to the primary constraint set $\mathcal{P}_{t, \zeta^{Y}} \subset T^{\ast}\mathscr{Y}_{t}$ by means of $\eqref{bspt12}$ (see~\cite{gotay1, molgado1} for more details).\\

We are now in a position to study the action of the Gauge symmetry group $\mathcal{G}$ on $\mathscr{Z}_{t}^{\star}$ and $T^{\ast}\mathscr{Y}_{t}$. We start by recalling that $\mathcal{G}$ acts on $Z^{\star}$ by means of canonical transformations, which give rise to an associated covariant momentum map $J^{(\mathcal{M})}: Z^{\star}\rightarrow \mathfrak{g}^{\ast} \otimes \bigwedge^{n-1}T^{\ast}Z^{\star}$. As discussed in~\cite{gotay1}, the latter induce the so-called energy-momentum map $E_{t}:\mathscr{Z}_{t}^{\star} \rightarrow \mathfrak{g}^{\ast}$ defined by 
\begin{equation}\label{bspt13}
    \langle E_{t}(\sigma), \xi \rangle:= \int_{\Sigma_{t}}\sigma^{\ast} J^{(\mathcal{M})}(\xi_{\eta}),
\end{equation}
where $\sigma \in \mathscr{Z}_{t}^{\star} $ and $\xi_{\eta}\in \mathfrak{g}$. In addition, to obtain a bona fide momentum map on $ \mathscr{Z}_{t}^{\star}$, we restrict our attention to the subgroup $\mathcal{G}_{t}:= \{\eta \in \mathcal{G} \left. | \right. \eta_{X}(\Sigma_{t}) = \Sigma_{t}\}$ of $\mathcal{G}$ that acts on $\Sigma_{t}$ by diffeomorphisms. Let $\mathfrak{g}_{t} \subset \mathfrak{g}$ be the Lie algebra of $\mathcal{G}_{t}$. Note that, for all $\eta \in \mathcal{G}_{t}$, the map $\eta_{t}:=\eta_{X} \left. | \right._{\Sigma_{t}}$ is in $\text{Diff}\,(\Sigma_{t})$. Then, the infinitesimal generators of $\text{Diff}\,(\Sigma_{t})$ will be denoted by $\xi_{\eta_{t}}^{X}$, and, in adapted coordinates, satisfy $\xi^{0}=0$ on $\Sigma_{t}$. It follows that, the energy-momentum map $E_{t}:\mathscr{Z}_{t}^{\star} \rightarrow \mathfrak{g}^{\ast}$ restricted to $\mathcal{G}_{t} \in \mathcal{G}$ gives rise to the momentum map $\mathcal{J}_{t}: E_{t} \left. | \right._{\mathfrak{g}_{t}}:\mathscr{Z}_{t}^{\star} \rightarrow \mathfrak{g}_{t}^{\ast} $, which corresponds to the momentum map for the action of $\mathcal{G}_{t}$ on $\mathfrak{g}_{t}$. In addition, since $(T^{\ast}\mathscr{Y}_{t}, \omega_{t})$ is the symplectic quotient of the presymplectic manifold $(\mathscr{Z}_{t}^{\star}, \Omega_{t})$ by the map $R_{t}: \mathscr{Z}_{t}^{\star} \rightarrow T^{\ast}\mathscr{Y}_{t}$ , we see that the momentum map $\mathscr{J}_{t}: T^{\ast}\mathscr{Y}_{t} \rightarrow \mathfrak{g}_{t}$ associated with the action of $\mathcal{G}_{t}$ on $T^{\ast}\mathscr{Y}_{t}$ is given by 
\begin{equation}\label{bspt14}
    \langle  \mathscr{J}_{t}(\varphi, \pi), \xi_{\eta}  \rangle:= \langle \mathcal{J}_{t}(\sigma), \xi_{\eta} \rangle,
\end{equation}
where $\xi_{\eta} \in \mathfrak{g}_{t}$, $(\varphi, \pi)\in T^{\ast}\mathscr{Y}_{t}$ and $\sigma \in R_{t}^{-1}\{(\varphi, \pi)\} \subset \mathscr{Z}_{t}^{\star} $.\\
Furthermore, since the first jet prolongation of $\zeta^{Y}$ to $J^{1}Y$ leaves the action $\eqref{la1}$ invariant, the corresponding lift of $\zeta^{Y}$ to $Z^{\star}$, $\zeta^{Z^{\star}} \in \mathfrak{X}(Z^{\star})$, acts on $Z^{\star}$ by covariant canonical transformations. Hence, the covariant momentum map associated with $\zeta^{Z^{\star}}$ projects onto a well defined function on the primary constraint set $\mathcal{P}_{t, \zeta^{Y}} \subset T^{\ast}\mathscr{Y}_{t}$, which, as proved in~\cite{gotay1}, coincides with the instantaneous Hamiltonian functional of the theory, $H_{t,\zeta^{Y}}: \mathcal{P}_{t, \zeta^{Y}} \rightarrow \mathbb{R}$, and can be defined by 
\begin{equation}\label{bspt15}
    H_{t,\zeta^{Y}}(\varphi, \pi):=-\int_{\Sigma_{t}} \sigma^{\ast}J^{(\mathcal{M})}(\zeta^{Z^{\star}}),
\end{equation}
where $\sigma \in R_{t}^{-1}\{ (\varphi, \pi) \} \cap \mathscr{N}_{t} $.\\
Finally, we want to point out here that for classical field theories with localizable symmetries, as explained in~\cite{gotay1, molgado1, molgado3}, the second Noether theorem within the multisymplectic formalism ensures that the zero level set of the momentum map $\eqref{bspt14}$ gives rise to the set of first-class constraints that characterizes the system within the instantaneous Dirac-Hamiltonian approach, namely, 
\begin{equation}
     \mathscr{J}_{t}^{-1}(0):= \{ (\varphi, \pi)\in T^{\ast}\mathscr{Y}_{t} \left. | \right.  \langle  \mathscr{J}_{t}(\varphi, \pi), \xi_{\eta}  \rangle:= 0, \forall \xi_{\eta} \in \mathfrak{g}_{t}\} 
\end{equation}
\section{Geometric-Covariant analysis of the Palatini-Cartan model for gravity} \label{sec3}
The purpose of this section is to study the Palatini-Cartan model for gravity within the geometric-covariant Lagrangian, multisymplectic and polysymplectic formulations for classical field theories. To this end, we will begin by introducing the mathematical description of the model under consideration, the Palatini-Cartan model for gravity, which is a four dimensional formulation for General Relativity with cosmological constant. Here, we will discuss how the different geometric-covariant formalism previously introduced generates new insights in the geometry underlying the model under study. We will deduce the field equations and also we will analyze the symmetries described by the action of the Gauge group over the background spacetime manifold. An important step in this section is the polysymplectic analysis of the model since, as we will see, the introduction of a Dirac-Poisson bracket structure results completely non-trivial. Moreover, after performing a space-time decomposition of the smooth manifold where the model is defined, we will recover the instantaneous Dirac-Hamiltonian analysis of the theory.

\subsection{Palatini-Cartan model for gravity}
As we mentioned before, the Palatini-Cartan model for gravity is a 4-dimensional formulation for the Einstein theory of General Relativity that includes a cosmological constant $\Lambda$.\footnote{From now on, the dimension of the background manifold is $n=4$ and we will fix the signature of the metric to $\sigma=1$} Here, we will do a mathematical description of the model of our interest. \\
To begin with, let us consider $X$ as a 4-dimensional Lorentzian manifold without boundary. Then, given $\mathfrak{g}$ a Lie algebra (or a vector subspace of a Lie algebra), we can identify $\Omega^{p}(X,\mathfrak{g}):=\Omega^{p}(X)\otimes \mathfrak{g}$ with the set of $\mathfrak{g}$-valued $p$-forms on $X$. Additionally, let $\{T_{a}\}$ be a basis for $\mathfrak{g}$ and $\beta:  \mathfrak{g} \otimes \mathfrak{g} \rightarrow \mathbb{R}$ a symmetric bilinear form. Thus, for any $\mu, \lambda \in \Omega^{p}(X, \mathfrak{g})$, we define $[ \mu, \lambda ]:= ( \mu^{a} \wedge \lambda^{b}) \otimes [T_{a}, T_{b}]$, $d\mu := d\mu^{a} \otimes T_{a}$ and $\beta(\mu \wedge \lambda):= \mu^{a}\wedge \lambda^{b}\, \beta(T_{a}, T_{b})$, where $[\cdot, \cdot]$ is the Lie bracket. Now, consider the group $ \mathcal{H}=SO(3,1)$ and $\mathfrak{h}:= \mathfrak{so} (3, 1)$ its corresponding Lie algebra. Thus, following~\cite{wise1} we can consider the Palatini model as a Cartan-type gauge theory by considering a Cartan connection $A=\omega + e$, where $\omega$ is a $\mathfrak{h}$-valued connection and $e$ is a co-frame field. With this in mind, we can think of the Palatini's model as a gauge theory with gauge group $\mathcal{G}$, whose Lie algebra is denoted by $\mathfrak{g}$, which may be identified with $ SO(4,1),\, SO(3,2),\, ISO(3,1)$ depending on whether $\Lambda$ is correspondingly positive, negative, or zero, as discussed in~\cite{wise1,wise2}.
Now, by regarding $\mathcal{H}$ as a closed subgroup of $\mathcal{G}$, we can perform the symmetric splitting $\mathfrak{g}= \mathfrak{h} \oplus \mathfrak{p}$ (for more details about symmetric splittings see~\cite{wise2, book1}), where the supplement space $\mathfrak{p}\subset \mathfrak{g}$ satisfies the commutation relations $\left[\mathfrak{h}, \mathfrak{p} \right] \subset \mathfrak{p}$ and $\left[\mathfrak{p}, \mathfrak{p} \right] \subset \mathfrak{h}$. It follows that if $\{J_{ab}\}$ spans $\mathfrak{h}$  and $\{P_{c}\}$ spans $\mathfrak{p}$, then $\{J_{ab}, P_{c}\}$ is a basis for $\mathfrak{g}$ such that 
\begin{align}
    \left[ J_{ab}, J_{cd} \right]&= - \left(\eta_{da}J_{cb} + \eta_{bd} J_{ac}- \eta_{ca}J_{db} - \eta_{bc}J_{ad}\right),\nonumber \\
\left[ J_{ab}, P_{c} \right]&=- \left(\eta_{ac} P_{b}- \eta_{bc} P_{a}\right),\nonumber \\
\left[ P_{a}, P_{b} \right] &=\frac{ \Lambda}{2} \epsilon_{ab}{}^{cd} J_{cd},
\end{align}
where $\epsilon_{abcd}$ is the 4-dimensional Levi Civita alternating symbol and $\eta_{ab}$ stands for the internal metric ($a,b,c,d \in \{0, 1, 2, 3\}$). \\
Furthermore, as pointed out in~\cite{wise1,wise2}, we can define a bilinear form $\beta: \mathfrak{g}\otimes \mathfrak{g} \rightarrow \mathbb{R}$, which, when restricted to $\mathfrak{h}$, can be written as 
\begin{equation} \label{C1}
    \beta|_{\mathfrak{h}}(J_{ab},J_{cd}):= \alpha\, Tr(J_{ab} \star (J_{cd}) ),
\end{equation}
where $\alpha:=\frac{3}{2\Lambda}$ is a weight factor and $\star$ is the internal Hodge operator on $\mathfrak{h}$. Note that, since $\alpha$ is a weight factor, the bilinear form $ \beta|_{\mathfrak{h}}$ is non-degenerate for all $\Lambda$. \\
In light of this, given $\eqref{C1}$ and thinking of the $\mathfrak{h}$-connection $\omega \in \Omega^{1}(M, \mathfrak{h})$ and the co-frame $e \in \Omega^{1}(M, \mathfrak{p})$ as part of a Cartan connection $A:= \omega + e \in \Omega^{1}(M, \mathfrak{g})$, the Palatini-Cartan model for gravity is written as 
\begin{equation}
    S_{P}(A):= \int_{X} \left. \beta \right|_{\mathfrak{h}}\left(F_{\mathfrak{h}} \wedge F_{\mathfrak{h}}- R  \wedge R\right),
\end{equation}
where $F \rightarrow F_{\mathfrak{h}}$ is the projection of the curvature of $A$ into $\mathfrak{h}$ given by 
\begin{equation}
F_{\mathfrak{h}}:= R \left[ \omega \right] - \frac{\Lambda}{3} e \wedge e,
\end{equation}
with $R[\omega]= d\omega + \frac{1}{2}\left[ \omega, \omega \right]$ denoting the $\mathfrak{h}$-valued curvature 2-form. Observe that, in terms of the $\mathfrak{h}$-connection and the co-frame field $e$, we can introduce the Palatini-Cartan action as 
\begin{equation} \label{C2}
S_{P}[e, \omega]= \int_{X} Tr\left( e \wedge e \wedge \star R \left[ \omega \right] - \frac{\Lambda}{6} e\wedge e \wedge \star\left( e \wedge e \right)  \right)
\end{equation}
An advantage of considering the Palatini-Cartan model for gravity as a Cartan-type gauge theory is that it defines a topological-like field theory which is invariant under both gauge transformations and diffeomorphism~\cite{molgado1, catren1}. However, the gauge transformations of the Cartan connection  $A \in \Omega^{1}(M, \mathfrak{g})$ can be generated through a combination of gauge transformations associated with the so-called $\mathcal{H}$-gauge and internal translational symmetries of the theory, which are related with the first-class constraints that emerge in the instantaneous Dirac-Hamiltonian analysis of the model~\cite{nikolic1}. We are therefore in a position to study the gauge symmetries of the system. In this regard, we find that the gauge transformations associated with the $\mathcal{H}$-gauge symmetry of the model read 
\begin{align}\label{gt1}
     e \rightarrow e_{\theta}&:= e + \left[ e, \theta \right],\nonumber\\
        \omega \rightarrow \omega_{\theta}&:= \omega + d_{\omega}\theta ,
\end{align}
where $d_{\omega}:= d + \left[ \omega, \cdot \right]$ is the covariant derivative associated with the $\mathfrak{h}$-connection $\omega$, and $\theta \in \Omega^{0}(X, \mathfrak{h})$ is an arbitrary $\mathfrak{h}$-valued function on $X$. Further, the gauge  transformations associated with the internal translational symmetry of the model are given by 
\begin{align}\label{gt2}
      e \rightarrow e_{\chi}&:= e + d_{\omega} \chi,\nonumber \\
        \omega \rightarrow \omega_{\chi}&:= \omega + \left[ e, \chi \right],
\end{align}
where $\chi \in \Omega^{0}(X, \mathfrak{p})$ is an arbitrary $\mathfrak{p}$-valued function on $X$. As discussed in~\cite{catren1, Gronwald, witten}, the co-frame field $e$  allows us identify a vector field generating internal gauge translations as the generator of external spacetime diffeomorphisms of $X$, then providing a correspondence between the internal gauge translations and the external spacetime diffeomorphisms of $X$. Of course, the extended gauge symmetry group of the Palatini-Cartan model for gravity corresponds to $\mathcal{G}$, such that the elements of $\mathfrak{h}$ give rise to the $\mathcal{H}$-gauge symmetry and the elements of $\mathfrak{p}$ are responsible for the generation of the internal translational symmetry.\\ 
Next, we will analyze within the covariant-Lagrangian formalism the features of the gauge and diffeomorphism invariant field theory $\eqref{C2}$ focusing our attention in analyzing the $\mathcal{H}$-gauge and translational symmetries of the system. 

\subsection{Lagrangian analysis}
In this subsection, we will analyze the Palatini-Cartan model for gravity from the point of view of the geometric-covariant Lagrangian formulation for classical field theories. Here, we will not only obtain the field equations, but we will also study the gauge symmetries of the theory, which will be fundamental to construct the associated Noether currents of the theory.\\
To start, consider $Y:=\left(T^{\ast}X \otimes \mathfrak{h} \right) \otimes \left( T^{\ast}X \otimes \mathfrak{p}  \right)$. Then, we can define $\left( Y, \pi_{XY}, X\right)$ as the configuration space of the Palatini-Cartan model for gravity. This is because the dynamical fields of such physical system can be understood as local sections of the bundle $\pi_{XY}$. In particular, given a local coordinate system $(x^{\mu})$ on $X$, we can identify $\left( x^{\mu}, a_{\mu}^{a}, b_{\mu}^{ab} \right)$ as an adapted local coordinate system on $Y$. It follows that given a local section $\phi \in \mathscr{Y}_{X}$ it can be locally represented by $\left( x^{\mu}, e_{\mu}^{a}, \omega_{\mu}^{ab} \right)$. Moreover, given $\left( J^{1}Y, \pi_{EJ^{1}Y},Y \right)$ the affine jet bundle over $Y$, we can induce a local adapted coordinate system on $J^{1}Y$ denoted by $\left(  x^{\mu}, a_{\mu}^{a}, b_{\mu}^{ab}, a_{\mu \nu}^{a}, b_{\mu \nu}^{ab}  \right)$. Then, as discussed in section~\ref{sec2}, the first jet prolongation  $j^{1}\phi \in \mathscr{J}{}^{1}\mathscr{Y}_{X} $ of a section $\phi \in \mathscr{Y}_{X}$, can be locally represented as $\left( x^{\mu}, e_{\mu}^{a}, \omega_{\mu}^{ab} , \partial_{\mu}e_{\nu}^{a},  \partial_{\mu}\omega_{\nu}^{ab}  \right)$. With this in mind, the action principle $\eqref{C2}$ can be rewritten as 
\begin{equation}\label{act1}
    S_{P}[e, \omega]= \int_{X} d^{n}x \left(j^{1}\phi\right)^{\ast}L_{P},
\end{equation}
with the Lagrangian function $L_{P}: J^{1}Y \rightarrow \mathbb{R}$ explicitly given by 
\begin{equation}
    L_{P}(x,a,b)= \frac{1}{4} \epsilon^{\mu \nu \rho \sigma} \epsilon_{abcd}\left(a_{\mu}^{a}a_{\nu}^{b}R_{\rho \sigma}^{cd} - \frac{\Lambda}{6}a_{\mu}^{a}a_{\nu}^{b} a_{\rho}^{c}a_{\sigma}^{d} \right)
\end{equation}
where $R_{\mu \nu}^{ab} := b_{\mu\nu}^{ab} - b_{\nu\mu}^{ab} + \epsilon^{ab}{}_{cd}b_{\mu}{}^{cl}b_{\nu}{}_{l}{}^{d}$ stands for the components of the curvature  $\mathfrak{h}$-valued 2-form $R$, and $\epsilon_{abcd}$ comes from the structure constants of $\mathfrak{h}$.\\
As mentioned before, within the geometry-covariant Lagrangian approach the Lagrangian function is not the main object of interest, but the so-called Poincaré-Cartan forms, $\Theta_{P}^{(\mathcal{L})} \in \Omega^{n}_{1} (J^{1}Y)$ and $\Omega_{P}^{(\mathcal{L})}:=-d\Theta_{P}^{(\mathcal{L})}   \in \Omega^{n+1}_{1} (J^{1}Y)$ which, by construction, contains all the dynamical information of the theory under study. In our case, the Poincaré-Cartan $n$-form is given by 
\begin{align}\label{La1}
    \Theta_{P}^{(\mathcal{L})}&=\frac{1}{2}  \epsilon^{\mu \nu \rho \sigma} \epsilon_{abcd} a_{\rho}^{a} a_{\sigma}^{b} db_{\nu}^{cd} \wedge d^{n-1}x_{\mu} \nonumber\\
&- \frac{1}{4} \epsilon^{\mu \nu \rho \sigma} \epsilon_{abcd} \left( 2a_{\mu}^{a} a_{\nu}^{b} b_{\rho \sigma}^{cd} - a_{\mu}^{a}a_{\nu}^{b}R_{\rho \sigma}^{cd} + \frac{\Lambda}{6}a_{\mu}^{a}a_{\nu}^{b} a_{\rho}^{c}a_{\sigma}^{d} \right)d^{n}x
\end{align}
As we will see below, the Poincaré-Cartan forms will be useful to obtain not only the field equations of the theory but also the conserved currents associated with the action of the gauge group over the configuration space of the model by means of the $\mathcal{H}$-gauge symmetry and the translational symmetry. \\
Now, in order to obtain the correct field equations of the theory let us consider an arbitrary vector field $X \in \mathfrak{X}(J^{1}Y)$ on $J^{1}Y$ given by 
\begin{equation}
X= X^{(1)}{}_{\mu}^{a} \frac{\partial}{\partial a_{\mu}^{a}} + X^{(2)}{}_{\mu}^{ab} \frac{\partial }{\partial b_{\mu}^{ab}}
\end{equation}
By direct calculation, it is not hard to see that 
\begin{align}
    X \lrcorner d\Theta_{P}^{(\mathcal{L})} =& \epsilon^{\mu \nu \alpha \beta} \epsilon_{abcd}  \left( X^{(1)}{}_{\mu}^{a}  a_{\nu}^{b}  \left( db_{\beta}^{cd}\wedge d^{n-1}x_{\alpha}- b_{\alpha \beta}^{cd}  d^{n}x + \frac{1}{2}  R_{\alpha \beta}^{cd} d^{n}x- \frac{\Lambda}{6}  a_{\alpha}^{c} a_{\beta}^{d} d^{n}x \right) \right.\nonumber \\
& \left. -  X^{(2)}{}_{\mu}^{cd}   a_{\nu}^{b} \left(  da_{\alpha}^{a} \wedge d^{n-1}x_{\beta} + b_{\beta}{}^{a}{}_{l} a_{\alpha}^{l} d^{n}x \right) \right)
\end{align}
Besides, given $\phi \in \mathscr{Y}_{X}$ a critical point for the action principle $\eqref{C2}$, the condition $\eqref{lat1}$ must hold, and hence we can write 
\begin{equation}
\epsilon^{\mu \nu \alpha \beta} \epsilon_{abcd} \left( X^{(1)}{}_{\mu}^{a}  e_{\nu}^{b} \left( R_{\alpha \beta}^{cd} - \frac{\Lambda}{3}  e_{\alpha}^{c} e_{\beta}^{d} \right) +  2X^{(2)}{}_{\mu}^{cd}   e_{\nu}^{b} D_{\alpha}e_{\beta}^{a}  \right)=0
\end{equation}
Note that since $X \in \mathfrak{X}(J^{1}Y)$ is an arbitrary vector field on $J^{1}
Y$, the above gives rise to the following set of equations
\begin{align}\label{fe1}
    R_{\alpha \beta}^{cd} - \frac{\Lambda}{6} \epsilon_{abcd}  e_{\alpha}^{c} e_{\beta}^{d} &=0, \nonumber\\
  D_{[\alpha}e_{\beta]}^{a}& =0,
\end{align}
where $D_{\alpha}e_{\beta}^{a};= \partial_{\alpha}e_{\beta}^{a} + \omega_{\alpha}{}^{a}{}_{c}\, e_{\beta}^{c}$ denotes the local representation of the covariant derivative $d_{\omega}$.
The above set of equations corresponds to the field equations of the Palatini-Cartan model for gravity, namely, the Einstein equations and the torsion-free condition, respectively~\cite{wise2, romano1, peldan1}.\\

Now, we will analyze the action of the extended gauge symmetry group on the configuration space at the Lagrangian level. To this end, as explained in section~\ref{sec2}, given $\xi_{\theta} \in \mathfrak{h}$ and $\xi_{\chi} \in \mathfrak{p}$, we start by defining $\xi_{\theta}^{Y} \in \mathfrak{X}(Y)$ and $\xi_{\chi}^{Y} \in \mathfrak{X}(Y)$ as infinitesimal generators of the gauge transformations $\eqref{gt1}$ and $\eqref{gt2}$, respectively. Locally, these vector fields reads 
\begin{align}\label{pla1}
    \xi_{\theta}^{Y}&= \epsilon^{a}{}_{bcd}a_{\mu}^{b}\theta^{cd} \frac{\partial }{\partial a_{\mu}^{a}} + D_{\mu}\theta^{ab}\frac{\partial }{\partial b_{\mu}^{ab}}\nonumber \\
         \xi_{\chi}^{Y}&= D_{\mu}\chi^{a}\frac{\partial }{\partial a_{\mu}^{a}}  + \frac{\Lambda}{2} \epsilon^{ab}{}_{cd}a_{\mu}^{c}\chi^{d} \frac{\partial }{\partial b_{\mu}^{ab}}
\end{align}
Furthermore, we know that the fiber-preserving transformations on the covariant configuration space of a given classical field theory induce a fiber-preserving transformation on the corresponding affine jet bundle. In this sense, by considering $\eqref{la2}$, the first jet  prolongations, $ \xi_{\theta}^{J^{1}Y} \in \mathfrak{X}(J^{1}Y)$ and $ \xi_{\chi}^{J^{1}Y} \in \mathfrak{X}(J^{1}Y)$, associated with the infinitesimal generators $\eqref{pla1}$ can be locally written as 
\begin{align}\label{vf1}
        \xi_{\theta}^{J^{1}Y} &= \epsilon^{a}{}_{bcd}a_{\mu}^{b}\theta^{cd} \frac{\partial }{\partial a_{\mu}^{a}} + D_{\mu}\theta^{ab}\frac{\partial }{\partial b_{\mu}^{ab}} + \epsilon^{a}{}_{bcd}\left( a_{\nu}^{b} \partial_{\mu} \theta^{cd} + a_{\mu \nu}^{b} \theta^{cd} \right) \frac{\partial }{\partial a_{\mu \nu}^{a}} \nonumber \\
		& \ \ \  + \left( D_{\nu}\partial_{\mu}\theta^{cd} + \epsilon^{ab}{}_{cd} b_{\mu \nu}^{ce} \theta_{e}{}^{d}\right) \frac{\partial }{\partial b_{\mu \nu}^{ab}},\nonumber\\
        \xi_{\chi}^{J^{1}Y} &= D_{\mu}\chi^{a}\frac{\partial }{\partial a_{\mu}^{a}}  + \frac{\Lambda}{2}  \epsilon^{ab}{}_{cd}a_{\mu}^{c}\chi^{d} \frac{\partial }{\partial b_{\mu}^{ab}} + D_{\nu}\partial_{\mu}\chi^{a} \frac{\partial }{\partial a_{\mu \nu}^{a}} + \frac{\Lambda}{2}  \epsilon^{ab}{}_{cd} a_{\nu}^{c} \partial_{\mu}\chi^{d}  \frac{\partial }{\partial b_{\mu \nu}^{ab}} .
\end{align}
Bearing this in mind, let us study the action of the gauge symmetries of the Palatini-Cartan model for gravity on its Poincare-Cartan $n$-form $\eqref{La1}$. To begin with, let us first consider the case of the $\mathcal{H}$-gauge symmetry. By direct calculation, we find that for all $\xi_{\theta} \in \mathfrak{h}$, this gauge symmetry preserves the Poincare-Cartan $n$-form up to an exact form, namely 
\begin{equation} \label{lsc1}
\mathcal{L}_{\xi_{\theta}^{J^{1}Y}}\Theta_{P}^{(\mathcal{L})}=d\alpha_{\theta}^{(\mathcal{L})},
\end{equation}
where $\alpha_{\theta}^{(\mathcal{L})} \in \Omega_{0}^{n-1}(J^{1}Y)$ is a $\pi_{YJ^{1}Y}$-horizontal $(n-1,0)$-form on $J^{1}Y$ given in coordinates by 
\begin{equation}\label{ns1}
	\alpha_{\theta}^{(\mathcal{L})}= 2 \epsilon^{\mu \nu \alpha \beta}   a_{\alpha}^{a} a_{\beta}^{b} b_{\nu}{}_{ac} \theta^{c}{}_{b}  d^{n-1}x_{\mu}.
    \end{equation}
Similarly, a straightforward calculation shows that, for all $\xi_{\chi} \in \mathfrak{p}$, the translational symmetry also preserves the Poincare-Cartan $n$-form up to an exact form, specifically 
\begin{equation} \label{lsc2}
\mathcal{L}_{\xi_{\chi}^{J^{1}Y}}\Theta_{P}^{(\mathcal{L})}=d\alpha_{\chi}^{(\mathcal{L})}
\end{equation}
being $\alpha_{\chi}^{(\mathcal{L})} \in \Omega_{0}^{n-1}(J^{1}Y) $ a $\pi_{YJ^{1}Y}$-horizontal $(n-1,0)$-form on $J^{1}Y$ given in coordinates by 
\begin{equation} \label{ns2}
    \alpha_{\chi}^{(\mathcal{L})}=  \epsilon^{\mu \nu \alpha \beta} \left[2a_{\nu}^{a}\left(\partial_{\alpha}\chi^{b} b_{\beta ab} - b_{\alpha ac} b_{\beta}{}^{c}{}_{b}\chi^{b}\right) - \frac{\Lambda}{6} \epsilon_{abcd} \chi^{a}a_{\nu}^{b}a_{\alpha}^{c}a_{\beta}^{d}\right] \, d^{n-1}x_{\mu}.
\end{equation}
Furthermore, we affirm that both the $\mathcal{H}$-gauge and the translational symmetries are localizable symmetries. To show this, let us consider the case of $\mathcal{H}$-gauge symmetry, since analogous arguments follow for translational symmetry. To start, note that for all $\theta \in \Omega^{0}(X, \mathfrak{h})$, both the vector field $\eqref{vf1}$ and the $\pi_{EJ^{1}E}$-horizontal $(n-1,0)$-form $\eqref{ns1}$ are linear on the components of  $\theta$ and their partial derivatives, then it follows that the set $\mathcal{C}_{LS}^{\mathcal{H}} := \{ (\xi_{\theta}^{J^{1}E}, \alpha_{\theta}^{(\mathcal{L})} )\}$ is a vector space. In addition, since the condition $\eqref{lsc1}$ holds, each $(\xi_{\theta}^{J^{1}E}, \alpha_{\theta}^{(L)})$ is a Noether symmetry. Finally, let $U_{1}, U_{2}$ be open sets of $X$ such that its closure is disjoint, and consider $\theta, \hat{\theta} \in \Omega^{0}(M. \mathfrak{h})$. Suppose that $\hat{\theta}= \theta$ on $U_{1}$ but $\hat{\theta}=0$ on $U_{2}$, then $\xi_{\theta}^{J^{1}E}(p)=\hat{\xi}_{\theta}^{J^{1}E}(p)$ and $\alpha_{\theta}^{(L)}(p)= \hat{\alpha}_{\theta}^{(L)}(p)$ for all $p \in \pi_{1}^{-1}(U_{1})$ but $\hat{\xi}_{\theta}^{J^{1}E}(p)=0$ and $ \hat{\alpha}_{\theta}^{(L)}(q)=0$ for all $q \in \pi_{1}^{-1}(U_{2})$. This shows that $\mathcal{C}_{LS}^{\mathcal{H}}:= \{(\xi_{\theta}^{J^{1}E}, \alpha_{\theta}^{(L)})\}$ is a set of localizable symmetries. \\

Thus, the action of $\mathcal{G}$ on $J^{1}Y$ has an associated Lagrangian covariant momentum map, namely $J^{(L)}: J^{1}Y \rightarrow \mathfrak{g}^{\ast} \otimes \bigwedge^{n-1} T^{\ast}J^{1}Y $. In particular, considering the splitting $\mathfrak{g}= \mathfrak{h} \oplus \mathfrak{p}$, we can obtain the local representation of such a map, namely, $J^{(L)}(\xi)$  for an arbitrary  $\xi \in \mathfrak{g}$, as follows. First, let $\xi_{\theta}\in \mathfrak{h}$ and $\xi_{\chi}\in \mathfrak{p}$. Then, by considering $\eqref{la8}$, and in light of the relations $\eqref{lsc1}$ and $\eqref{lsc2}$ we can write 
\begin{align}\label{lmm1}
    J^{(\mathcal{L})}\left(\xi_{\theta}\right)&= \epsilon^{\mu \nu \alpha \beta} a_{\alpha}^{a} a_{\beta}^{b} \left(  \partial_{\nu} \theta_{ab}- b_{\nu}{}_{a}{}^{c}\, \theta_{cb} \right)d^{n-1}x_{\mu}, \nonumber \\
    J^{(\mathcal{L})}\left(\xi_{\chi}\right)&=- \epsilon^{\mu \nu \alpha \beta} \left[2a_{\nu}^{a}\left(\partial_{\alpha}\chi^{b} b_{\beta ab} - b_{\alpha ac} b_{\beta}{}^{c}{}_{b}\chi^{b}\right) - \frac{\Lambda}{6} \epsilon_{abcd} \chi^{a}a_{\nu}^{b}a_{\alpha}^{c}a_{\beta}^{d}\right] \, d^{n-1}x_{\mu}.
\end{align}
Of course, the relation above is important because it will be useful to construct the Noether currents of the classical field theory under consideration by pulling back the Lagrangian covariant momentum map with a solution of the Euler-Lagrange fields equations of the theory. In this case, for all $\xi_{\theta} \in \mathfrak{h}$, $\xi_{\chi} \in \mathfrak{p}$ and $\phi \in \mathscr{Y}_{X}$ a solution for the field equations $\eqref{fe1}$, the $(n-1)$-forms on $X$ defined by 
\begin{align}\label{lmm2}
    \mathcal{J}^{(\mathcal{L})}\left(\xi_{\theta}\right)&= \epsilon^{\mu \nu \alpha \beta} e_{\alpha}^{a} e_{\beta}^{b} \left(  \partial_{\nu} \theta_{ab}- \omega_{\nu}{}_{a}{}^{c}\, \theta_{cb} \right)d^{n-1}x_{\mu}, \nonumber \\
    \mathcal{J}^{(\mathcal{L})}\left(\xi_{\chi}\right)&=-  \epsilon^{\mu \nu \alpha \beta} \left[2e_{\nu}^{a}\left(\partial_{\alpha}\chi^{b} \omega_{\beta ab} - \omega_{\alpha ac} \omega_{\beta}{}^{c}{}_{b}\chi^{b}\right) - \frac{\Lambda}{6} \epsilon_{abcd} \chi^{a}e_{\nu}^{b}e_{\alpha}^{c}e_{\beta}^{d}\right] \, d^{n-1}x_{\mu}.
\end{align}
correspond to the Noether currents associated with the $\mathcal{H}$-gauge and internal translational symmetries of the Palatini-Cartan model for gravity, respectively. Since the second Noether current in $\eqref{lmm2}$ is associated with internal gauge translations, it generates the external spacetime diffeomorphisms of $X$ on-shell. \\
Finally, note that by imposing the field equations $\eqref{fe1}$, the integral of this Noether current over a Cauchy surface $\Sigma_{t}\subset X$ vanishes, which is consistent with the fact that the symmetries of the theory are localizable, and therefore from the second Noether theorem follows that for all $\phi \in \mathscr{Y}_{X}$, a solution of the field equations of the theory, the Lagrangian Noether currents of the system must vanish on-shell. \\

In the following subsection, we will analyze the four dimensional gravity model $\eqref{C2}$ within the multisymplectic approach, focusing on the study of the gauge symmetries of the theory.

\subsection{Multisymplectic analysis }
In the present subsection, we will analyze the Palatini-Cartan model for gravity within the multisymplectic approach, focusing our attention on studying the action of the extended gauge group over its corresponding covariant multimomenta phase-space. Here, we will see how the $\mathcal{H}$-gauge and internal translational symmetries of the system give rise to covariant canonical transformations, which in turn will allow us to obtain the covariant momentum map associated with the extended gauge group of the theory. \\

To begin, let $\left( Z^{\star}, \pi_{Y Z^{\star} }, Y \right)$ be the covariant multimomenta phase space associated with the theory. As we saw before, given $\left( x^{\mu}, a_{\mu}^{a}, b_{\mu}^{ab} \right)$ an adapted local coordinate system on $Y$, the corresponding adapted coordinate system on $Z^{\star}$ is defined by $\left(  x^{\mu}, a_{\mu}^{a}, b_{\mu}^{ab}, p, p_{a}^{\mu \nu}, \bar{p}_{ab}^{\mu \nu} \right)$. Therefore, the canonical and multisymplectic forms of the Palatini-Cartan model for gravity, $\Theta_{P}^{(\mathcal{M})} \in \Omega^{n}\left(  Z^{\star} \right)$ and $\Omega_{P}^{(\mathcal{M})} \in \Omega^{n+1}\left(  Z^{\star} \right)$, respectively, can be locally written as 
\begin{align}
    \Theta_{P}^{(\mathcal{M})}&= p_{a}^{\mu \nu} da_{\nu}^{a}\wedge d^{n-1}x_{\mu} + \bar{p}_{ab}^{\mu \nu} db_{\nu}^{ab}\wedge d^{n-1}x_{\mu} + p d^{n}x,\nonumber\\
    \Omega_{P}^{(\mathcal{M})}&= da_{\nu}^{a}\wedge dp_{a}^{\mu \nu}\wedge d^{n-1}x_{\mu} + da_{\nu}^{a}\wedge d\bar{p}_{ab}^{\mu \nu} \wedge d^{n-1}x_{\mu}- dp \wedge d^{n}x.
\end{align}
Now, in order to study the $\mathcal{H}$-gauge and translational symmetries of the theory at the multisymplectic level, given $\xi_{\theta} \in \mathfrak{h}$ and $\xi_{\chi}\in \mathfrak{p}$, we start by identifying $\xi_{\theta}^{\theta} \in \mathfrak{X}^{1}(Z^{\star})$ and $\xi_{\theta}^{\chi} \in \mathfrak{X}^{1}(Z^{\star})$, that is , the $\alpha^{(\mathcal{M})}$-lifts of the vector fields $\eqref{ps1}$ to $Z^{\star}$, as the infinitesimal generators of the gauge transformations $\eqref{gt1}$ and $\eqref{gt2}$ on the covariant multimomenta phase space, respectively. These vector fields are given locally by 
\begin{align}\label{lm1}
    \xi_{\theta}^{\alpha}&=  \epsilon^{a}{}_{bcd}a_{\mu}^{b}\theta^{cd} \frac{\partial }{\partial a_{\mu}^{a}} + D_{\mu}\theta^{ab}\frac{\partial }{\partial b_{\mu}^{ab}} - \epsilon^{aecd} \bar{p}_{cd}^{\mu \nu} \theta_{e}{}^{b} \frac{\partial}{\partial \bar{p}_{ab}^{\mu \nu} } +  \epsilon^{ab}{}_{cd}\,p_{b}^{\mu \nu} \theta^{cd} \frac{\partial}{\partial p_{a}^{\mu \nu} } \nonumber\\
      &  \ \ \ \  - \left( \epsilon^{a}{}_{bcd}\, p_{a}^{\mu \nu} a_{\nu}^{b} \partial_{\mu}\theta^{cd} + \bar{p}_{ab}^{\mu \nu} D_{\nu}\partial_{\mu}\theta^{ab} \right) \frac{\partial }{\partial p},\nonumber \\
     \xi_{\chi}^{\alpha}&= D_{\mu}\chi^{a}\frac{\partial }{\partial a_{\mu}^{a}}  + \frac{\Lambda}{2} \epsilon^{ab}{}_{cd}a_{\mu}^{c}\chi^{d} \frac{\partial }{\partial b_{\mu}^{ab}} -  \left(  p_{a}^{\mu \nu} D_{\nu}\partial_{\mu}\chi^{a} + \epsilon^{ab}{}_{cd}\, \bar{p}_{ab}^{\mu \nu} a_{\nu}^{c} \partial_{\mu}\chi^{d}\right) \frac{\partial }{\partial p}.
\end{align}
With this in mind, we have that, for all $\xi_{\theta}\in \mathfrak{h}$, the $\mathcal{H}$-gauge symmetry of the system acts on $Z^{\star}$ through covariant canonical transformations, namely, 
\begin{equation}\label{lscm1}
\mathcal{L}_{\xi_{\theta}^{\alpha}}\Theta_{P}^{(\mathcal{M})}=d\alpha_{\theta}^{(\mathcal{M})},
\end{equation}
where $\alpha_{\theta}^{(\mathcal{M})} \in \Omega_{0}^{n-1}(Z^{\star})$ is a $\pi_{EZ^{\star}}$-horizontal $(n-1,0)$-form on $Z^{\star}$ given in coordinates by 
\begin{equation}\label{nsm1}
	\alpha_{\theta}^{(\mathcal{M})}= 2 \epsilon^{\mu \nu \alpha \beta}   a_{\alpha}^{a} a_{\beta}^{b} b_{\nu}{}_{ac} \theta^{c}{}_{b}  d^{n-1}x_{\mu}.
\end{equation}
Similarly, for all $\xi_{\chi} \in \mathfrak{p}$, the translational symmetry also acts on $Z^{\star}$ through covariant canonical transformations as the following relation holds 
\begin{equation} \label{lscm2}
\mathcal{L}_{\xi_{\chi}^{Z^{\star}}}\Theta_{P}^{(\mathcal{M})}=d\alpha_{\chi}^{(\mathcal{M})},
\end{equation}
where $\alpha_{\chi}^{(\mathcal{M})} \in \Omega_{0}^{n-1}(Z^{\star}) $ is a $\pi_{EZ^{\star}}$-horizontal $(n-1,0)$-form on $Z^{\star}$ given in coordinates by 
\begin{equation} \label{ns2}
    \alpha_{\chi}^{(\mathcal{M})}= \epsilon^{\mu \nu \alpha \beta} \left[2a_{\nu}^{a}\left(\partial_{\alpha}\chi^{b} b_{\beta ab} - b_{\alpha ac} b_{\beta}{}^{c}{}_{b}\chi^{b}\right) - \frac{\Lambda}{6} \epsilon_{abcd} \chi^{a}a_{\nu}^{b}a_{\alpha}^{c}a_{\beta}^{d}\right] \, d^{n-1}x_{\mu}.
\end{equation}
Thus, the action of the gauge group $\mathcal{G}$ on $Z^{\star}$ induces a covariant momentum map $J^{(\mathcal{M})}: Z^{\star}\rightarrow \mathfrak{g}^{\ast}\otimes \bigwedge^{n-1} T^{\ast}Z^{\star}$. In fact, as in the Lagrangian case, the splitting $\mathfrak{g}= \mathfrak{h}\oplus \mathfrak{p}$ allows us to obtain a local representation of the covariant momentum map, namely, $J^{(\mathcal{M})}(\xi)\in \Omega^{n-1}(Z^{\star})$ for all $\xi \in \mathfrak{g}$. To this end, let $\xi_{\theta} \in \mathfrak{h}$ and $\xi_{\chi}\in \mathfrak{p}$ then in light of relations $\eqref{lscm1}$ and $\eqref{lscm2}$, it is possible to write 
\begin{align}\label{cmm1}
    J^{(\mathcal{M})}(\xi_{\theta})&=\left[ \epsilon^{a}{}_{bcd} \, p^{\mu \nu}_{a} a_{\nu}^{b} \theta^{cd} + \bar{p}^{\mu \nu}_{ab}D_{\nu}\theta^{ab} - 2 \epsilon^{\mu \nu \alpha \beta}   a_{\alpha}^{a} a_{\beta}^{b} b_{\nu}{}_{ac} \theta^{c}{}_{b}  \right] d^{n-1}x_{\mu},\nonumber \\
         J^{(\mathcal{M})}(\xi_{\chi})&= \left[ p_{a}^{\mu \nu} D_{\nu}\chi^{a} + \frac{\Lambda}{2}\epsilon^{ab}{}_{cd} \bar{p}_{ab}^{\mu\nu} a_{\nu}^{c} \chi^{d} - \epsilon^{\mu \nu \alpha \beta} \biggl(2a_{\nu}^{a}\left(\partial_{\alpha}\chi^{b} b_{\beta ab} - b_{\alpha ac} b_{\beta}{}^{c}{}_{b}\chi^{b}\right)  \right. \nonumber\\
          &  \ \ \ \left. - \frac{\Lambda}{6} \epsilon_{abcd} \chi^{a}a_{\nu}^{b}a_{\alpha}^{c}a_{\beta}^{d}\biggr)  \right]d^{n-1}x_{\mu}.
\end{align}
It is not difficult to see that we can recover the Lagrangian momentum maps $\eqref{lmm1}$ simply by pulling back the covariant momentum maps $\eqref{cmm1}$ with the Legendre transformation $\eqref{m8}$. Finally, we would like to mention that, as we will see in the subsequent subsections, the covariant momentum maps $\eqref{cmm1}$ associated with the extended gauge group of the model, obtained at the multisymplectic level, will be fundamental to recover, on the space of Cauchy data, the set of first-class constraints of the theory within the instantaneous Dirac-Hamiltonian formulation and, also, will allow us to construct conserved currents for the solutions of the De Donder-Weyl-Hamilton field equations. \\

Next, we will carry out the polysymplectic formulation for the gravity model $\eqref{C2}$, where we will emphasize the relevance of the Dirac-brackets within the treatment of the second class constraints of the system.  

\subsection{Polysymplectic analysis}
In this subsection, we will analyze the Palatini-Cartan model for gravity from the point of view of the polysymplectic formalism. As it is well known, the Palatini-Cartan model for gravity corresponds to a singular Lagrangian system, however, as we mentioned in the introduction of this article, in the multisymplectic formalism there is not yet a well-defined Poisson bracket, then we can not perform a Dirac-Hamiltonian-like analysis for this singular model within the multisymplectic approach but, as we explained in subsection~\ref{ssp}, there is a well-defined Poisson bracket within the polysymplectic formalism. In consequence, we will make use of the algorithm described in subsection~\ref{ssp} to study this kind of systems within the polysymplectic approach. Here, we will obtain the correct De Donder-Weyl-Hamilton field equations of the theory and the conserved currents induced by the covariant momentum map associated with the extended gauge group of the system. \\
To begin with, let us consider the quotient bundle $\left(P, \pi_{YP}, Y \right)$. Then, given 
$\left( x^{\mu}, a_{\mu}^{a}, b_{\mu}^{ab} \right)$ an adapted local coordinate system on $Y$, we denote by $\left( x^{\mu}, a_{\mu}^{a}, b_{\mu}^{ab}, p_{a}^{\mu \nu}, \bar{p}_{ab}^{\mu \nu} \right)$ an adapted coordinate system on $P$. Thus, a section $\varrho \in \mathscr{P}_{X}$ of $\pi_{XP}$, the polymomenta phase space of the theory,  can be locally represented as $\left( x^{\mu}, a_{\mu}^{a}, b_{\mu}^{ab}, \pi_{a}^{\mu \nu}, \bar{\pi}_{ab}^{\mu \nu} \right)$. \\
Now, in order to describe the Palatini model for gravity in a covariant Hamiltonian-like formulation, we apply the covariant Legendre transformation, which allows us to write 
\begin{align}\label{pa1}
    p_{a}^{\mu \nu}&:= \frac{\partial L_{P}}{\partial a_{\mu\nu}^{a}}=0, \nonumber \\
     \bar{p}_{ab}^{\mu \nu}&:=\frac{\partial L_{P}}{\partial b_{\mu\nu}^{ab}}=\frac{1}{2} \epsilon^{\mu \nu \alpha \beta}\epsilon_{abcd} a_{\alpha}^{c} a_{\beta}^{d}.
\end{align}
Now, the relation $\eqref{pa1}$ gives rise to the primary constraints of the Palatini-Cartan model for gravity which can be identified with $(n-1, 0)$-forms explicitly as 
\begin{align}\label{pa2}
    C^{(1)}{}_{a}^{\nu}&:= p_{a}^{\mu \nu} \bar{\omega}_{\mu}, \nonumber  \\
   C^{(2)}{}_{ab}^{\nu}&:= \left(\bar{p}_{ab}^{\mu \nu} - \frac{1}{2} \epsilon^{\mu \nu \alpha \beta}\epsilon_{abcd} a_{\alpha}^{c} a_{\beta}^{d} \right)\bar{\omega}_{\mu}. 
\end{align}
Furthermore, we can define a primary constraint surface $P_{\mathrm{Poly}}^{PCS} \subset P$ characterized by the vanishing of the primary constraint $(n-1)$-forms $\eqref{pa2}$. Moreover, we have that on $P_{\mathrm{Poly}}^{PCS} \subset P$ the polymomenta variables are antisymmetric in the space-time indices, and hence we are able to define a more suitable set of canonically conjugate variables for which the commutation relations under the Poisson-Gerstenhaber bracket $\eqref{ps8}$ are given by 
\begin{align}\label{pa3}
    \left\lbrace \left[ p_{a}^{\mu \nu} \bar{\omega}_{\mu}, a_{[\sigma}^{b} \bar{\omega}_{\rho ]} \right] \right\rbrace &=\delta_{a}^{b}\delta_{[\sigma}^{\nu} \bar{\omega}_{\rho]},\nonumber\\
\left\lbrace  \left[ \bar{p}_{ab}^{\mu \nu}\bar{\omega}_{\mu}, b_{[\sigma}^{cd} \bar{\omega}_{\rho]} \right]\right\rbrace &= \delta_{a}^{c}\delta_{b}^{d}\delta_{[\sigma}^{\nu}\bar{\omega}_{\rho]}.
\end{align}
Note that, this choice of canonically conjugate variables for the bracket structure associated with the physical model under consideration is related to the fact that the dynamical information of a gauge theory whose fields are Lie algebra-valued 1-forms is contained in the antisymmetric part of the polymomenta variables, as discussed in~\cite{ deleon1, book1}. Thus, the set of canonically conjugate variables introduced in $\eqref{pa3}$ will allow us to consistently describe the system under study within the polysymplectic approach, as we will see throughout this subsection. \\
With this in mind, it is not hard to see that, by considering $\eqref{ps4}$, the De Donder-Weyl-Hamiltonian associated with the Palatini model for gravity can be written as \begin{equation}\label{pa4}
    H_{DW}=-\frac{1}{2} \epsilon^{\mu \nu \alpha\beta} \epsilon_{abcd}\left( a_{\mu}^{a} a_{\nu}^{b} b_{\alpha}{}^{c}{}_{e} b_{\beta}{}^{ed} - \frac{\Lambda}{12} a_{\mu}^{a}a_{\nu}^{b} a_{\alpha}^{c}a_{\beta}^{d}   \right).
\end{equation} 
However, since the Palatini-Cartan model for gravity corresponds to a singular system, it is well known that the De Donder-Weyl-Hamilton field equations reproduced by $\eqref{pa4}$ might not be equivalent to the Euler-Lagrange field equations of the physical model, as described by the Dirac treatment for constrained systems~\cite{henb1}.  \\
Thus, in order to develop a correct polysymplectic analysis for the gravity model $\eqref{C2}$, we will use the algorithm to study singular Lagrangian systems within the polysymplectic formalism. To this end, we begin by introducing the total De Donder-Weyl-Hamiltonian of the system, namely 
\begin{equation}
    \bar{H}_{DW}:= H_{DW} + \lambda_{(1)}{}_{\nu}^{a}\bullet C^{(1)}{}^{\nu}_{a} + \lambda_{(2)}{}^{ab}_{\nu} \bullet C^{(2)}{}^{\nu}_{ab},
\end{equation}
where $ \lambda_{(1)}{}_{\nu}^{a}:= \lambda_{(1)}{}_{\mu \nu}^{a}dx^{\mu} \in \Omega^{1}_{0}(P)$ and $ \lambda_{(2)}{}_{\nu}^{ab}:=  \lambda_{(2)}{}_{\mu \nu}^{ab}dx^{\mu} \in \Omega^{1}_{0}(P)$ denotes a set of Lagrange multiplier $(1;0)$-forms enforcing the primary constraint $(n-1)$-forms $\eqref{pa2}$, and $\bullet$ is the co-exterior product defined by $\eqref{ps7}$.
Now, to verify whether the set $\eqref{pa2}$ of $(n-1)$-constraints of the theory is complete, we need to impose the consistency condition on each of its elements. Hence, using the polysymplectic structure 
\begin{equation}\label{pos1}
    \Omega_{P}^{(P)}= dp^{\mu \nu}_{a}\wedge da_{\nu}^{a} \wedge \bar{\omega}_{\mu} + d\bar{p}^{\mu \nu}_{ab}\wedge db_{\nu}^{ab} \wedge \bar{\omega}_{\mu},
\end{equation}
it is not difficult to see that the consistency condition applied to the first $(n-1)$-constraint in $\eqref{pa2}$ leads to the following relation 
\begin{equation}\label{pa5}
    d\bullet C^{(1)}{}^{\mu}_{a}=  -\epsilon^{\mu \nu \alpha \beta}\epsilon_{abcd} \left( a_{\nu}^{b} b_{\alpha}{}^{c}{}_{e} b_{\beta}{}^{ed} + \frac{\Lambda}{6}  a_{\nu}^{b} a_{\alpha}^{c}a_{\beta}^{d} \right) + \epsilon^{\mu \nu \alpha \beta} \epsilon_{abcd}  \lambda_{(2)}{}_{\nu \alpha}^{bc} a_{\beta}^{d} \approx 0,
\end{equation}
whereas a similar calculation shows that the consistency condition associated with the second $(n-1)$-constraint in $\eqref{pa2}$ gives rise to the relation
\begin{equation}\label{pa6}
 d\bullet C^{(2)}{}^{\mu}_{ab}= \epsilon^{\mu \nu \alpha \beta}\epsilon_{abcd} b_{\nu}{}^{c}{}_{e} a_{\alpha}^{e} a_{\beta}^{d} + \epsilon^{\mu \nu \alpha \beta}\epsilon_{abcd}  \lambda_{(1)}{}_{\nu \alpha}^{c} a_{\beta}^{d} \approx 0. 
\end{equation}
Note that both equations $\eqref{pa5}$ and $\eqref{pa6}$ impose restrictions on the Lagrange multipliers $ \lambda_{(2)}{}_{\mu \nu}^{ab} \in \Omega^{1}_{0}(P) $ and $\lambda_{(1)}{}_{\mu \nu}^{a} \in \Omega^{1}_{0}(P)$, respectively, which implies that there are no secondary constraints, and therefore the set of constraints $(n-1)$-forms $\eqref{pa2}$ of the Palatini-Cartan model is complete. Explicitly, the consistency conditions allow us to fix the components of the Lagrange multipliers in the following way 
\begin{align}
    \lambda_{(1)}{}_{[\mu \nu]}^{a}&= - b_{[\mu}{}^{a}{}_{c}a_{\nu]}^{c}, \nonumber \\
\lambda_{(2)}{}_{[\mu \nu]}^{ab}&= - \frac{1}{2} \epsilon^{ab}{}_{cd} \left( b_{\mu}{}^{c}{}_{e} b_{\nu}{}^{ed} - \frac{\Lambda}{6}a_{\mu}^{c} a_{\nu}^{d}\right). 
\end{align} 
Consequently, we are now in position to classify the set of constraints $(n-1)$-forms of the model. To this end, note that the commutation relations of the constraints $(n-1)$-forms $\eqref{pa2}$ under the Poisson-Gerstenhaber bracket are given by 
\begin{align}\label{pa7}
    \{ [ C^{(1)}{}_{a}^{\mu},   C^{(2)}{}_{bc}^{\nu} ] \} &:=\epsilon^{\sigma \mu \nu  \gamma} \epsilon_{abcd}a_{\gamma}^{d} \bar{\omega}_{\sigma},\nonumber\\
\{ [ C^{(1)}{}_{a}^{\mu},  C^{(1)}{}_{b}^{\nu} ] \} &:=0,\nonumber\\ 
\{ [ C^{(2)}{}_{ab}^{\mu},  C^{(2)}{}_{cd}^{\nu}] \} &:=0.
\end{align}
Now, the $(n-1, 0)$-form matrix valued $C^{(i,j)}:=\{ [ C^{(i)},  C^{(j)} ] \}$ defined by $\eqref{pa7}$ can be therefore written as 
\begin{equation}\label{pa8}
C^{(i,j)}:=\left(
\begin{matrix}
0 & C^{(1,2)}{}_{abc}^{\mu \nu} \\
C^{(2,1)}{}_{abc}^{\mu \nu}& 0
\end{matrix} \right),
\end{equation}
where $C^{(1,2)}{}_{abc}^{\mu \nu}:= \{ [ C^{(1)}{}^{\mu}_{a},  C^{(2)}{}^{\nu}_{bc}] \}$ and $C^{(2,1)}{}_{abc}^{\mu \nu}:=\{ [ C^{(2)}{}^{\mu}_{ab},  C^{(1)}{}^{\nu}_{c}] \}$ are rectangular block matrices ($16 \times 24$ in $n=4$ dimensions). Then, since the matrix $\eqref{pa8}$ does not vanish on the surface $P_{\mathrm{Poly}}^{PCS} \subset P$,  it follows that the set of constraints $(n-1)$-forms $\eqref{pa2}$ is complete and also second-class, in Dirac's terminology, unlike what occurs in the standard Dirac-Hamiltonian analysis for Palatini-Cartan gravity, in which, as described in~\cite{nikolic1}, the set of constraints of the theory is characterized by second-class primary and first-class secondary constraints. However, one reason why the sets of constraints in these formalisms are not the same is that the Poisson structure within the polysymplectic approach is defined by the Poisson-Gerstenhaber bracket $\eqref{ps8}$, rather than the usual Poisson bracket used in the standard Dirac-Hamiltonian formalism~\cite{henb1}. Moreover, in the usual Dirac's Hamiltonian formulation dealing with a matrix like $\eqref{pa8}$ would mean that the number of second-class constraints of the system is incorrect, since $C^{(i,j)}$ in $\eqref{pa8}$ could be singular, implying the existence of first-class constraints, but the case is substantially different here since within the polysysmplectic approach the number of polymomenta is different from the number of field variables, as detailed in~\cite{kana1}. \\
With this in mind, our main objective now is to construct the Dirac-Poisson bracket for the Palatini-Cartan model for gravity. However, since the matrix $\eqref{pa8}$ is rectangular, we can not use an ordinary inverse matrix to construct a well-defined Dirac-Poisson bracket $\eqref{ps21}$. A well-known generalized inverse for arbitrary matrices is the so-called generalized Moore-Penrose inverse, for more details about the Moore-Penrose inverse see~\cite{penrose1, mpi1}. 
As we shall see, the Dirac-Poisson bracket constructed with this Moore-Penrose-type generalized inverse will allow us to eliminate the second class constraints $\eqref{pa2}$, and eventually obtain the correct De Donder-Weyl field equations of the system. To this end, it should be noted that, since the constraints of the theory are organized in Hamiltonian $(n-1)$-forms, the Dirac-Poisson bracket must be defined for dynamical variables given by Hamiltonian $0$- or $(n-1)$-forms~\cite{kana4}. Thus, let $F \in \Omega^{0}_{0}(P) $ or $F \in  \Omega^{n-1}_{0}(P)$ and $G \in \Omega^{n-1}_{0}(P)$ be Hamiltonian forms. Then in our case, the Dirac-Poisson bracket $\eqref{ps21}$ reduces to
\begin{equation}\label{pa9}
	\{ [ F,  G ] \}_{D}:= \{ [ F,  G ] \} - \{[ F,  C^{(i)} ] \} \bullet \left( C^{-1}{}_{(i,j)} \wedge \{ [ C^{(j)},  G ] \}  \right)
\end{equation}
where $C^{-1}{}_{(i,j)}$ stands for the Moore-Penrose-type generalized inverse $(1; 0)$-form valued matrix such that~\cite{kana1, penrose1}
\begin{equation}\label{pa10}
C^{(i,j)} \bullet \left(  C^{-1}{}_{(j,k)} \wedge C^{(k,l)} \right)= C^{(i,l)},
\end{equation}
where the Latin indices $i,j,k,...$ run over the complete set of constraint $(n-1)$-forms $\eqref{pa2}$. Note that the specific block structure of $\eqref{pa8}$ ensures that the  generalized Moore-Penrose inverse of $C^{(i,j)}$ has the same block structure as $\eqref{pa8}$ with its components, respectively, replaced by $ C^{-1}{}_{(1,2)}{}^{abc}_{\mu \nu}$ and $C^{-1}{}_{(2,1)}{}^{abc}_{\mu \nu}$, that is,
\begin{equation}\label{pa12}
C^{-1}_{(i,j)}:=
\left(
\begin{matrix}
0 & C^{-1}{}_{(1,2)}{}^{abc}_{\mu \nu} \\
C^{-1}{}_{(2,1)}{}^{abc}_{\mu \nu}& 0
\end{matrix} \right).
\end{equation}
Furthermore, without having an explicit expression for $\eqref{pa12}$ we can obtain a useful relation between its inverse components, as using $\eqref{pa9}$ we get
\begin{equation}
    \begin{split}
        \{[a_{\mu}^{a}, b_{[ \nu}^{bc}\bar{\omega}_{\sigma]}]\}_{D}&=C_{(1,2)}^{-1}{}_{\mu\nu}^{abc},\\
         \{[ b_{[ \nu}^{bc}\bar{\omega}_{\sigma]}, a_{\mu}^{a}]\}_{D}&=C_{(2,1)}^{-1}{}_{\nu\mu}^{bca}.
    \end{split} 
\end{equation}
But since the Dirac-Poisson bracket must fulfill the properties of a graded Lie algebra, it follows that 
\begin{equation}\label{pa11}
    C_{(1,2)}^{-1}{}_{\mu\nu}^{abc}=-C_{(2,1)}^{-1}{}_{\nu\mu}^{bca}
\end{equation}
In particular, for each component of $\eqref{pa8}$ the relations $\eqref{pa10}$ must hold, and then its inverse $(1,0)$-forms are explicitly given by 
\begin{equation}\label{pa12}
\begin{split}
C^{-1}_{(1,2)}{}^{abc}_{\mu \nu}&=-\frac{1}{3!\cdot 4}\epsilon^{abcd}\epsilon_{\sigma \mu \nu \gamma} a_{d}^{\gamma}dx^{\sigma},\\
C^{-1}_{(2,1)}{}^{abc}_{\mu \nu}&=\frac{1}{3!\cdot4}\epsilon^{abcd}\epsilon_{\sigma \mu \nu \gamma} a_{d}^{\gamma}dx^{\sigma},
\end{split}
\end{equation}
where we had used $\eqref{pa11}$. As a consequence, condition $\eqref{ps20}$ is fulfilled in the subspace generated by the co-frame $e \in \Omega^{1}(M, \mathfrak{p})$, namely
\begin{equation}
    C_{(1,k)}^{-1}{}_{\alpha \beta}^{abc} \wedge C^{(k, 1)}{}_{bcd}^{\beta \gamma}:= \delta_{a}^{d}\delta_{\alpha}^{\gamma}\bar{\omega}
\end{equation}
Now, with this Dirac-Poisson bracket $\eqref{pa9}$ at hand, which is compatible with  relation $\eqref{pa10}$, we can satisfactory study the Palatini-Cartan model for gravity on the constraint surface $P_{\mathrm{Poly}}^{PCS} \subset P$. To this end, a straightforward calculation shows that the commutation relations between the conjugate field variables $\eqref{pa3}$ under the Dirac-Poisson bracket $\eqref{pa9}$ are given by 
\begin{equation}
    \begin{split}
        \{ [ p_{a}^{\mu \nu}\bar{\omega}_{\mu},  a^{b}_{[\sigma}  \bar{\omega}_{\rho]}]\}_{D}&=0,\\
        \{ [ \bar{p}_{ab}^{\mu \nu}\bar{\omega}_{\mu},  b^{cd}_{[\sigma}  \bar{\omega}_{\rho]}]\}_{D}&= \delta^{c}_{a}\delta^{d}_{b} \delta_{[\sigma}^{\nu}\bar{\omega}_{\rho]}.
    \end{split}
\end{equation}
Additionally, a similar calculation shows that the useful commutation relation between the field variables is given by
\begin{equation}
        \{[a_{\mu}^{a}, b_{[\nu}^{cd} \bar{\omega}_{\sigma]}]\}_{D}=- \frac{1}{ 3! \cdot 4}\epsilon^{abcd}\epsilon_{\mu \nu \sigma \gamma} a_{d}^{\gamma},
\end{equation}
With this in mind, we are in a position to study the De Donder-Weyl-Hamilton field equations of the theory. Note that, on the primary constraint surface $P_{P}^{PCS} \subset P$, we can obtain the correct field equations of the model by means of the set of canonically conjugate variables $\eqref{pa3}$, the De Donder-Weyl-Hamiltonian $\eqref{pa4}$ and the Poisson bracket $\eqref{pa9}$. In other words, let $\varrho \in \mathscr{P}_{X}$ a section of $\pi_{XP}$, then the De Donder-Weyl-Hamilton field equations of the Palatini-Cartan model for gravity are given by 
\begin{subequations}\label{pa13}
\begin{align}
    \partial_{[\mu}e_{\nu]}^{a}&=\varrho^{\ast} \{[H_{DW}, a_{[\nu}^{a} \bar{\omega}_{\mu]}]\}_{D}=-\omega_{[\mu}{}^{a}{}_{b}e_{\nu]}^{b},\label{pa13a}\\
    \partial_{\mu}\pi_{a}^{\mu \nu}&=\varrho^{\ast} \{[H_{DW}, p_{a}^{\mu \nu} \bar{\omega}_{\mu]}]\}_{D}=0,\label{pa13b}\\
    \partial_{[\mu}\omega_{\nu]}^{ab}&=\varrho^{\ast} \{[H_{DW}, b_{[\nu}^{ab} \bar{\omega}_{\mu]}]\}_{D}=-\frac{1}{2} \epsilon^{ab}{}_{cd}\left( \omega_{\mu}{}^{c}{}_{e}\omega_{\nu}{}^{ed} - \frac{\Lambda}{6} e_{\mu}^{c} e_{\nu}^{d}\right),\label{pa13c} \\
    \partial_{\mu}\bar{\pi}_{a}^{\mu \nu}&=\varrho^{\ast} \{[H_{DW}, \bar{p}_{a}^{\mu \nu} \bar{\omega}_{\mu]}]\}_{D}=-2\epsilon_{ab}{}^{cd}e^{\mu}_{c} \omega_{[\mu}{}_{de}e_{\nu]}^{e}.\label{pa13d}
\end{align}
\end{subequations}
It is not hard to see that equation $\eqref{pa13a}$ can be explicitly written as 
\begin{equation}\label{pa14}
    D_{[\mu}e_{\nu]}^{a}=0,
\end{equation}
which is the torsion free condition for gravity. Similarly, equation $\eqref{pa13c}$ gives rise to the Einstein field equations 
\begin{equation}\label{pa15}
    R_{\mu \nu}^{ab}- \frac{\Lambda}{6} \epsilon^{ab}{}_{cd} e_{\mu}^{c} e_{\nu}^{d}=0.
\end{equation}
Also, note that in light of equation $\eqref{pa13a}$, the De Donder-Weyl-Hamilton field equation $\eqref{pa13b}$ and $\eqref{pa13d}$ are trivially satisfied. Therefore, we had derived the torsion free condition $\eqref{pa14}$ and the Einstein field equations $\eqref{pa15}$ from the point of view of the geometric-Lagrangian analysis of the Palatini-Cartan model for gravity, which shows the equivalence between the Lagrangian and the De Donder-Weyl-Hamilton field equations. \\

To finish this section, note that we can induce the covariant momentum map associated with the Palatini-Cartan model for gravity within the polymomenta phase space. To this end, having the splitting $\mathfrak{g}= \mathfrak{h} \oplus \mathfrak{p}$, let $\xi_{\theta}\in \mathfrak{h}$ and $\xi_{\chi}\in \mathfrak{p}$ and $h_{DW}\in \mathscr{Z}^{\star}_{P} $ be the De Donder-Weyl-Hamiltonian section of the bundle $\pi_{PZ^{\star}}: Z^{\star}  \rightarrow P$, then we can obtain a local representation of the covariant momentum map on the polymomenta phase space, that is $J^{(P)}(\xi)=h_{DW}^{\ast} J^{(\mathcal{M})}(\xi) \in \Omega_{0}^{n-1}(P)$ for all $\xi \in \mathfrak{g}$, which can be written in terms of $(n-1)$-forms on $P$ as 
\begin{align}
    J^{(\mathcal{P})}(\xi_{\theta})&:=\left[ \epsilon^{a}{}_{bcd} \, p^{\mu \nu}_{a} a_{\nu}^{b} \theta^{cd} + \bar{p}^{\mu \nu}_{ab}D_{\nu}\theta^{ab} - 2 \epsilon^{\mu \nu \alpha \beta}   a_{\alpha}^{a} a_{\beta}^{b} b_{\nu}{}_{ac} \theta^{c}{}_{b}  \right] \bar{\omega}_{\mu}, \nonumber \\
         J^{(\mathcal{P})}(\xi_{\chi})&:= \left[ p_{a}^{\mu \nu} D_{\nu}\chi^{a} + \frac{\Lambda}{2}\epsilon^{ab}{}_{cd} \bar{p}_{ab}^{\mu\nu} a_{\nu}^{c} \chi^{d} - \epsilon^{\mu \nu \alpha \beta} \biggl( 2a_{\nu}^{a}\left(\partial_{\alpha}\chi^{b} b_{\beta ab} - b_{\alpha ac} b_{\beta}{}^{c}{}_{b}\chi^{b}\right)  \right.\\
          &  \left. - \frac{\Lambda}{6} \epsilon_{abcd} \chi^{a}a_{\nu}^{b}a_{\alpha}^{c}a_{\beta}^{d} \biggr)  \right]d^{n-1}x_{\mu}.
\end{align}
Note that, as in the Lagrangian approach, given $\varrho \in \mathscr{P}_{X}$ a section of $\pi_{XP}$ the integration of the conserved current $\mathcal{J}^{(P)}(\xi)= \varrho^{\ast}J^{(P)}(\xi)$, on a Cauchy surface $\Sigma_{t} \subset X$ vanishes, and thus, $Q_{\Sigma_{t}}^{\mathcal{P}} (\xi)=0$, for all $\xi \in \mathfrak{g}$.

\subsection{The Palatini-Cartan model for gravity in the space of Cauchy data}
In this last section, we will carry out the space plus time decomposition for the Palatini-Cartan model for gravity $\eqref{C2}$.  Our main objective here is to describe how both the first-class constraints and the extended Hamiltonian of the theory can be recovered by means of geometric objects defined on the multimomenta phase space of the system. \\
To start with, let $\Sigma_{t} \subset X$ be a Cauchy surface of $X$, characterized by a level set of the local coordinate $x^{0}$, specifically $x^{0}-t=0$, for some $t \in \mathbb{R}$. We identify $\zeta^{X}= \partial_{0}$ as the infinitesimal generator of the slicing of the manifold $X$. Thus, for some $\xi_{\theta} \in \mathfrak{h}$ and $\xi_{\chi} \in \mathfrak{p}$, we introduce $(\Sigma_{t}, \zeta^{X})$ and $(Y_{t}, \zeta^{Y})$ to be the $\mathcal{G}$-slicing of the covariant configuration space $\pi_{XY}$, with the vector field $\zeta^{Y} \in \mathfrak{X}(Y)$ defined as  
\begin{equation}\label{spt1}
    \zeta^{Y} := \partial_{0} + \xi_{\theta}^{Y} + \xi_{\chi}^{Y},
\end{equation}
where $\xi_{\theta}^{Y} \in \mathfrak{X}(Y)$ and $\xi_{\chi}^{Y} \in \mathfrak{X}(Y)$ are explicitly given in coordinates by identities $\eqref{pla1}$. In what follows, we will understand $ \zeta^{Y} $ as a temporal direction on $Y$ for the Palatini-Cartan model for gravity. \\
Now, let $(Y_{t}, \pi_{\Sigma_{t}Y_{t}}, \Sigma_{t})$ be the restriction of the bundle $\pi_{XY}$ to $\Sigma_{t}$, then we can identify $(x^{i}, a_{\mu}^{a}, b_{\mu}^{ab})$, with $i \in \{ 1,2 ,3, \dotsc \}$, as an adapted coordinate system on $Y_{t}$. Thus, given $\mathscr{Y}_{t}$ the set of sections of $\pi_{\Sigma_{t}Y_{t}}$ we identify $T\mathscr{Y}_{t}$ as the $t$-instantaneous space of the velocities of the theory. Moreover, note that $\varphi=\phi \circ \imath_{t} $, for some $\phi \in \mathscr{Y}_{X}$, which allows us to identify $(e_{\mu}^{a}, \omega_{\mu}^{ab}, \dot{e}_{\mu}^{a},\dot{\omega}_{\mu}^{ab})$ as an adapted coordinate on $T\mathscr{Y}_{t}$, where the temporal derivative of the field variables reads
\begin{align}\label{spt2}
    \dot{e}_{\mu}^{a}&= \partial_{0}e_{\mu}^{a}-\epsilon^{a}{}_{bcd}e_{\mu}^{b}\theta^{cd}- D_{\mu}\chi^{a},\nonumber \\
        \dot{\omega}_{\mu}^{ab}&= \partial_{0}\omega_{\mu}^{ab} - D_{\mu}\theta^{ab} - \frac{\Lambda}{2}\epsilon^{ab}{}_{cd}e_{\mu}^{c}\chi^{d},
\end{align}
respectively. With this at hand, we can proceed to perform the space plus time decomposition of the Palatini-Cartan model for gravity $\eqref{C2}$ at the Lagrangian level. For this purpose, let $((J^{1}Y)_{t}, \pi_{Y_{t}(J^{1}Y)_{t}}, Y_{t})$ be the restriction of the bundle $\pi_{YJ^{1}Y}$ to $Y_{t}$. Then, the jet decomposition $\beta_{\zeta^{Y}}: (J^{1}Y)_{t} \rightarrow J^{1}Y_{t} \times VY_{t}$ over $Y_{t}$ is locally written as
\begin{equation}
    \beta_{\zeta^{Y}} (x^{i}, a_{\mu}^{a}, b_{\mu}^{ab}, a_{\mu\nu}^{a}, b_{\mu \nu}^{ab})=(x^{i}, a_{\mu}^{a}, b_{\mu}^{ab}, a_{i \nu}^{a}, b_{i \nu}^{ab},  \dot{e}_{\mu}^{a},\dot{\omega}_{\mu}^{ab} ) 
\end{equation}
Thus, it is not difficult to see that the instantaneous Lagrangian functional of the theory $L_{t, \zeta^{Y}}^{P} :T\mathscr{Y}_{t} \rightarrow \mathbb{R} $ reads 
\begin{align}\label{spt3}
    L_{t, \zeta^{Y}}^{P}(e, \omega, \dot{e}, \dot{\omega})&:=\frac{1}{2} \int_{\Sigma_{t}}d^{n-1}x_{0}\, \epsilon^{0ijk}\epsilon_{abcd}\biggl\{ e_{j}^{a}e_{k}^{b} \left(\dot{\omega}_{i}^{cd}  + \epsilon^{cd}{}_{mn}e_{i}^{m}\chi^{n} + D_{i}\theta^{cd}\right) + \nonumber \\
   & \left. 2\omega_{0}^{cd} e_{j}^{a} \left( \partial_{i}e_{k}^{b} - 
   \omega_{i}{}^{b}{}_{e} e_{k}^{e}\right)+ e_{0}^{a} e_{i}^{b} \left(  R_{jk}^{cd} - \frac{\Lambda}{3} e_{j}^{c} e_{k}^{d}\right) \right\},
\end{align}
where we performed some integration by parts and avoided terms on the boundary of $\Sigma_{t}$.\\
In particular, in light of the instantaneous Legendre transformation $\mathbb{F}\mathcal{L}_{t, \zeta^{Y}}^{P}: T\mathscr{Y}_{t} \rightarrow T^{\ast}\mathscr{Y}_{t} $ locally given by 
\begin{equation}
    \mathbb{F}\mathcal{L}_{t, \zeta^{Y}}^{P}(e, \omega, \dot{e}, \dot{\omega}):= (e^{a}_{\mu}, \omega_{\mu}^{ab}, \pi_{a}^{\mu\nu}, \bar{\pi}_{ab}^{\mu\nu}),
\end{equation}
the instantaneous momentum variables are given explicitly by 
\begin{align}\label{spt4}
    \pi_{a}^{\mu}&:= \frac{\partial  L_{t, \zeta^{Y}}^{P}}{\partial \dot{e}_{\mu}^{a}}=0, \nonumber \\
        \bar{\pi}_{ab}^{\mu}&:= \frac{\partial  L_{t, \zeta^{Y}}^{P}}{\partial \dot{\omega}_{\mu}^{ab}}= \frac{1}{2} \delta_{i}^{\mu} \epsilon^{0ijk}\epsilon_{abcd} e_{j}^{a}e_{k}^{b}.
\end{align}
Consequently, relations $\eqref{spt4}$ gives rise to the primary constraint surface $P_{t, \zeta^{Y}} \subset T^{\ast}\mathscr{Y}_{t}$ of the Palatini-Cartan model for gravity, that is, 
\begin{equation} \label{spt5}
    P_{t, \zeta^{Y}} := \{(e, \omega, \pi, \bar{\pi})\in  T^{\ast}\mathscr{Y}_{t}|\gamma_{a}^{\mu}=0, \varUpsilon_{a}^{i}=0, \bar{\gamma}_{ab}^{0}=0, \bar{\varUpsilon}_{ab}^{i}=0 \},
\end{equation}
where the primary constraints are defined by 
\begin{subequations}\label{spt6}
\begin{align}
     \gamma_{a}^{0}&:= \pi_{a}^{0}\approx 0, \label{spt6a}\\
        \varUpsilon_{a}^{i}&:=\pi_{a}^{i}\approx 0,\label{spt6b}\\
        \bar{\gamma}_{ab}^{0}&:= \bar{\pi}_{ab}^{0} \approx 0,\label{spt6c}\\
        \bar{\varUpsilon}_{ab}^{i}&:=   \bar{\pi}_{ab}^{i} - \frac{1}{2}  \epsilon^{0ijk}\epsilon_{abcd} e_{j}^{a}e_{k}^{b} \approx 0.\label{spt6d}
\end{align}
\end{subequations}
Of course, the presence of primary constraints on a classical field theory means that the  instantaneous Legendre transformation of the system is not an isomorphism. Thus, the Palatini-Cartan model for gravity $\eqref{C2}$ describes a Lagrangian singular system. \\
To continue, we will perform the space plus time decomposition of the Palatini-Cartan model at the  multisymplectic level. To start, let $(Z_{t}^{\star}, \pi_{Y_{t}Z_{t}^{\star}}, Y_{t})$ be the restriction of the bundle $\pi_{YZ}$ to $ Y_{t}$. In addition, denote $\mathscr{Z}_{t}^{\star}$ as the set of sections of the bundle $\pi_{\Sigma_{t}Z_{t}^{\star}}=\pi_{\Sigma_{t}Y_{t}} \circ \pi_{Y_{t}Z_{t}^{\star}}$, which is related to $T^{\ast}\mathscr{Y}_{t}$ through the bundle map $\mathcal{R}_{t}: \mathscr{Z}_{t}^{\star} \rightarrow T^{\ast}\mathscr{Y}_{t} $ over $\mathscr{Y}_{t}$. Furthermore, we identify $\zeta^{Z^{\star}} \in \mathfrak{X}(Z^{\star})$ as the $\alpha^{(\mathcal{M})}$-lift of $\eqref{spt1}$ to $Z^{\star}$, that is, 
\begin{equation} \label{spt10}
	\zeta^{Z^{\star}}= \partial_{0} + \xi_{\theta}^{\alpha} + \xi_{\chi}^{\alpha},
\end{equation}
where the vector fields $\xi_{\theta}^{\alpha} \in \mathfrak{X}(Z^{\star})$ and $\xi_{\chi}^{\alpha} \in \mathfrak{X}(Z^{\star})$ are given by $\eqref{lm1}$.\\
As already mentioned, we are interested in describing how the covariant momentum map associated with the gauge symmetry group $\mathcal{G}$ of the Palatini-Cartan model allows us to obtain the complete set of first-class constraints of the system. Note that, the gauge symmetry group of the theory $\mathcal{G}$ acts on $\pi_{XY}$ by means of a $\pi_{XY}$-bundle automorphism, which implies that the action of $\mathcal{G}$ on $ T^{\ast} \mathscr{Y}_{t}$ is well-defined, and hence there exists an associated covariant momentum map $ \mathscr{J}_{t}: T^{\ast}\mathscr{Y}_{t} \rightarrow \mathfrak{g}^{\ast}$. Then, to obtain a local representation of such a momentum map, we can proceed as follows. First, in light of the splitting $\mathfrak{g}=\mathfrak{h} \oplus \mathfrak{p}$, for all $\xi_{\eta} \in \mathfrak{g}$ its representation is given by $\xi = \xi_{\theta} + \xi_{\chi}$. Secondly, let $\sigma \in \mathscr{Z}_{t}^{\star} $ be a section such that $\mathcal{R}_{t}(\sigma):= (e, \omega, \pi, \bar{\pi}) \in  T^{\ast}\mathscr{Y}_{t}$. Hence, the induced momentum map associated with the action of $\mathcal{G}$ on $T^{\ast}\mathscr{Y}_{t}$ is given locally by 
\begin{equation}\label{spt11}
    \langle  \mathscr{J}_{t}(e, \omega, \dot{e}, \dot{\omega}), \xi_{\eta} \rangle=\int_{\Sigma_{t}}\sigma ^{\ast}J^{(\mathcal{M})}(\xi_{\eta})
\end{equation}
where $J^{(\mathcal{M})}(\xi_{\eta})= J^{(\mathcal{M})}(\xi_{\theta}) + J^{(\mathcal{M})}(\xi_{\chi}) $. To be precise, by using the covariant momentum maps $\eqref{cmm1}$ and denoting $\pi_{a}^{\mu}:= p_{a}^{0\nu} \circ \sigma$, $\bar{\pi}_{ab}^{\mu}:= \bar{p}_{ab}^{0\nu} \circ \sigma$, we find that  
\begin{align}\label{spt12}
    \langle  \mathscr{J}_{t}(e, \omega, \dot{e}, \dot{\omega}), \xi_{\eta} \rangle =& \int_{\Sigma_{t}}  d^{n-1}x_{0}  \left\lbrace  \left(D_{0}\chi^{a} + \epsilon^{a}{}_{bcd}e_{0}^{b} \theta^{cd} \right)\pi_{a}^{0} + \left( D_{0}\theta^{ab} + \frac{\Lambda}{2} \epsilon^{ab}{}_{cd} e_{0}^{c}\chi^{d} \right) \bar{\pi}_{ab}^{0} + \right. \nonumber \\
	& \left.   \left(D_{i}\chi^{a} + \epsilon^{a}{}_{bcd}e_{i}^{b} \theta^{cd} \right) \pi_{a}^{i} + \left( D_{i}\theta^{ab} + \frac{\Lambda}{2} \epsilon^{ab}{}_{cd} e_{i}^{c}\chi^{d} \right) \bar{\pi}_{ab}^{i}  - 2\epsilon^{0 ijk} a_{i}^{a}b_{j\, ac} a_{k}^{b}\theta^{c}{}_{b} \right. \nonumber\\
	& \left. -\epsilon^{0 ijk} \left[2a_{i}^{a}\left(\partial_{j}\chi^{b} b_{k ab} - b_{j \, ac} b_{k}{}^{c}{}_{b}\chi^{b}\right) - \frac{\Lambda}{6} \epsilon_{abcd} \chi^{a}a_{i}^{b}a_{j}^{c}a_{k}^{d}\right] \right\rbrace
\end{align}
Note that, in terms of the primary constraints $\eqref{spt6}$, the momentum map can be written as 
\begin{align}\label{spt13}
    \langle  & \mathscr{J}_{t}(e, \omega, \dot{e}, \dot{\omega}), \xi_{\eta} \rangle = \int_{\Sigma_{t}}  d^{n-1}x_{0} \biggl\{ \lambda_{a}^{0} \gamma_{a}^{0} + \bar{\lambda}_{0}^{ab}\bar{\gamma}_{ab}^{0} + \lambda^{a} \biggl[D_{i}\varUpsilon_{a}^{i} + \epsilon^{a}{}_{bcd} e_{b}{}_{i}\bar{\varUpsilon}_{cd}^{i} \nonumber \\
&  \left. + \epsilon^{0ijk} \left( R_{ij}{}_{ab} - \frac{\Lambda}{6}\epsilon_{abcd} e_{i}^{c} e_{j}^{d} \right) e_{k}^{b}  \right] + \bar{\lambda}^{ab}  \left(  \epsilon_{ab}{}^{cd} \varUpsilon_{c}^{i}e_{i}{}_{d} - D_{i}\bar{\varUpsilon}_{ab}^{i} +  \epsilon^{0ijk} D_{i}e_{j}{}_{a} \, e_{k}{}_{b} \right) \biggr\}  ,
\end{align}
where we avoided boundary terms and, also, we introduced the following definitions 
\begin{align}\label{spt14}
    \lambda_{0}^{a}&:= D_{0}\chi^{a} + \epsilon^{a}{}_{bcd}e_{0}^{b} \theta^{cd},\nonumber\\
	\lambda^{a}&=-\chi^{a}, \nonumber\\
	\bar{\lambda}_{0}^{ab}&:=D_{0}\theta^{ab} + \frac{\Lambda}{2} \epsilon^{ab}{}_{cd} e_{0}^{c}\xi^{d},\nonumber\\ 
	\bar{\lambda}^{ab}&:= \theta^{ab}.
\end{align}
Further, if we define
\begin{align}\label{spt15}
    \Gamma_{a}&:= D_{i}\varUpsilon_{a}^{i} + \epsilon^{a}{}_{bcd} e_{b}{}_{i}\bar{\varUpsilon}_{cd}^{i} +  \epsilon^{0ijk} \left( R_{ij}{}_{ab} - \frac{\Lambda}{6} \epsilon_{abcd} e_{i}^{c} e_{j}^{d} \right) e_{k}^{b}, \nonumber\\
	\bar{\Gamma}_{ab}&:=  \epsilon_{ab}{}^{cd} \varUpsilon_{c}^{i}e_{i}{}_{d} - D_{i}\bar{\varUpsilon}_{ab}^{i} + \epsilon^{0ijk} D_{i}e_{j}{}_{a} \, e_{k}{}_{b} ,
\end{align}
then we can rewrite $\eqref{spt11}$ in a simpler form as
\begin{equation}\label{spt16}
\langle   \mathscr{J}_{t}(e, \omega, \dot{e}, \dot{\omega}), \xi_{\eta} \rangle = \int_{\Sigma_{t}}  d^{n-1}x_{0}  \left\{ \lambda_{a}^{0} \gamma_{a}^{0} + \bar{\lambda}_{0}^{ab}\bar{\gamma}_{ab}^{0} + \lambda^{a}\Gamma_{a} + \bar{\lambda}^{ab}\bar{\Gamma}_{ab} \right\}.
\end{equation}
This last representation coincides with the generator of infinitesimal gauge transformations of the Palatini-Cartan model for gravity within Dirac's formulation~\cite{nikolic1}. \\
Moreover, it is important to remember that the gauge symmetries of the Palatini-Cartan model for gravity $\eqref{C2}$ correspond to localizable symmetries, then the admissible Cauchy data for the evolution equation of the theory is given by the zero level set of the momentum map $\eqref{spt11}$, namely 
\begin{align}\label{spt17}
    \mathscr{J}_{P\,t}^{-1}(0)&:= \{  (e, \omega, \dot{e}, \dot{\omega})\in T^{\ast} \mathscr{Y}_{t} | \langle   \mathscr{J}_{t}(e, \omega, \dot{e}, \dot{\omega}), \xi_{\eta} \rangle =0, \, \forall \xi_{\eta} \in \mathfrak{g}  \}\nonumber \\
	 &=\{  (e, \omega, \dot{e}, \dot{\omega})\in T^{\ast} \mathscr{Y}_{t} | \gamma_{a}^{0}=0, \bar{\gamma}_{ab}^{0}=0, \Gamma_{a}=0, \bar{\Gamma}_{ab}=0  \},
\end{align}
which is the surface of the $t$-instantaneous phase space defined by the set of first-class constraints characterizing the Palatini-Cartan model for gravity within the instantaneous Dirac-Hamiltonian formalism. In other worlds, the vanishing of the momentum map $\eqref{spt11}$ gives rise to the set of first-class constraints of the Palatini-Cartan model for gravity, namely, 
\begin{subequations} \label{spt18}
\begin{align}
	\gamma_{a}^{0} & \approx 0, \label{spt18a}\\
	 \bar{\gamma}_{ab}^{0} & \approx 0, \label{spt18b}\\
	 \Gamma_{a}& \approx 0,\label{spt18c} \\
	 \bar{\Gamma}_{ab} & \approx 0. \label{spt18d}
\end{align}
\end{subequations}
Note that, in Dirac's terminology, the primary constraints $\eqref{spt6b}$ and $\eqref{spt6d}$ which are not elements of the set $\eqref{spt18}$ must be second-class constraints. Moreover, if we consider the second-class primary constraints $\eqref{spt6b}$ and $\eqref{spt6d}$ as strong identities, the first-class constraints $\eqref{spt18c}$ and $\eqref{spt18d}$ reduce to 
\begin{align}\label{spt19}
    \Phi_{a}&:=   \epsilon^{0ijk} \left( R_{ij}{}_{ab} - \frac{\Lambda}{6} \epsilon_{abcd} e_{i}^{c} e_{j}^{d} \right) e_{k}^{b} \approx 0,\nonumber \\
	\Phi_{ab}&:= \epsilon^{0ijk} D_{i}e_{j}{}_{a} \, e_{k}{}_{b} \approx 0, 
\end{align}
which are the secondary constraints that arise when we impose the consistency conditions on the primary constraints. As discussed in~\cite{canepa1,canepa2, cattaneo1}, these constraints on $\Sigma_{t}$ are also first-class. The first one generates diffeormorphism along $\Sigma_{t}$, while stands for the torsion-free condition.\\
With all this in mind, we are in position to recover the $t$-instantaneous extended Hamiltonian function of the Palatini-Cartan model for gravity. To this end, let $\xi_{\theta} \in \mathfrak{h}$ and $\xi_{\chi} \in \mathfrak{p}$, then the vector field defined in $\eqref{spt10}$ satisfy $\mathcal{L}_{\zeta^{Z^{\star}}} \Omega_{P}^{(\mathcal{M})}=0$ and, hence, there is an $(n-1)$-form $H_{\zeta^{Z^{\star}}}^{(\mathcal{M})} := \zeta^{Z^{\star}} \lrcorner \Theta_{P}^{(\mathcal{M})}- \alpha_{\theta}^{(\mathcal{M})}-\alpha_{\chi}^{(\mathcal{M})} \in \Omega^{n-1}(Z^{\star})$ on $Z^{\star}$ such that $Z^{\star} \lrcorner \Theta_{P}^{(\mathcal{M})} = dH_{\zeta^{Z^{\star}}}^{(\mathcal{M})}$, where $\alpha_{\theta}^{(\mathcal{M})} \in \Omega^{n-1}(Z^{\star})$ and $\alpha_{\chi}^{(\mathcal{M})} \in \Omega^{n-1}(Z^{\star})$ are given in $\eqref{ns1}$ and $\eqref{ns2}$, respectively.  Now, for some $\phi \in \mathscr{Y}_{t}$, let us consider $\sigma \in \mathcal{N}_{t}= \{\sigma \in \mathscr{Y}_{t} | \sigma= \mathbb{F}\mathcal{L} \circ j^{1}\phi \circ \imath_{t} \}$ the canonical lift of an element $(e, \omega, \pi, \bar{\pi})\in P_{t, \zeta^{Y}}$. Then the instantaneous Hamiltonian functional $H_{t, \zeta^{Y}}:P_{t, \zeta^{Y}} \rightarrow \mathbb{R} $ of the system can be obtained by 
\begin{equation} 
	H_{t,\zeta^{Y}}(e, \omega, \pi, \bar{\pi}):= - \int_{\Sigma_{t}} \sigma^{\ast} H_{\zeta^{\star}}^{(\mathcal{M})}.
\end{equation}
Then, identifying $\pi_{a}^{\mu} := p_{a}^{0 \mu} \circ \sigma$ and $\bar{\pi}_{ab}^{\mu}:=\bar{p}_{ab}^{0 \mu} \circ \sigma$ we find that 
\begin{align}
    H_{t,\zeta^{Y}}(e, \omega, \pi, \bar{\pi})&=  \int_{\Sigma_{t}}d^{n-1}x_{0}\biggl\{\dot{e}_{0}^{a} \gamma_{a}^{0}  + \dot{\omega}_{0}^{ab} \bar{\gamma}_{ab}^{0} + \partial_{0}e_{i}^{a} \Upsilon_{a}^{i} + \partial_{0}\omega_{i}^{ab} \bar{\Upsilon}_{ab}^{i} + \chi^{a}\Gamma_{a} + \theta^{ab}\bar{\Gamma}_{ab}  \nonumber \\
& \left. -\epsilon^{0ijk}\epsilon_{abcd} \omega_{0}^{cd} e_{j}^{a} \left( \partial_{i}e_{k}^{b} - 
   \omega_{i}{}^{b}{}_{e} e_{k}^{e}\right)- \frac{1}{2}\epsilon^{0ijk}\epsilon_{abcd}e_{0}^{a} e_{i}^{b} \left(  R_{jk}^{cd} - \frac{\Lambda}{3} e_{j}^{c} e_{k}^{d}\right) \right\}
\end{align}
where we avoided some boundary terms of the Cauchy surface $\Sigma_{t}$. Moreover, under gauge fixing $e_{0}^{a} \approx 0$ and $\omega_{0}^{ab}\approx 0$, and considering the primary constraints as strong identities, we find  
\begin{equation}
    H_{t,\zeta^{Y}}(e, \omega, \pi, \bar{\pi}):= \int_{\Sigma_{t}}d^{n-1}x_{0}\left\lbrace  \chi^{a}\Phi_{a} + \theta^{ab}\bar{\Phi}_{ab}  \right\rbrace \approx 0,
\end{equation}
which is the extended $t$-instantaneous Hamiltonian in a reduced phase space presented in~\cite{canepa1,canepa2}.

\section{Conclusions}\label{sec4}
We reviewed the Palaniti-Cartan model for gravity within several covariant-geometric formulations for classical field theory.  At the covariant Lagrangian level, after introducing the Poincare-Cartan forms which describe the Lagrangian system, we recover the well-known field equations of the theory, namely the torsion free condition and the Einstein field equations. Moreover, the description of the Palatini model for gravity in terms of Cartan geometry allowed us to study its invariance under gauge transformations and spacetime diffeomorphisms in terms  of the $\mathcal{H}$-gauge symmetry and the translational symmetry, respectively. For each of these symmetries, we have constructed, from a pure geometrical perspective, the corresponding Lagrangian covariant momentum map, whose local representation contains the corresponding conserved quantities of the system. To the best of our knowledge, this latter construction had not  previously been explored for the Palatini-Cartan gravitational model within the covariant-geometric Lagrangian formulation for classical field theory. Besides, after constructing the multisymplectic form that characterizes the multimomentum phase space of the Palatini-Cartan model for gravity, we analyzed the action of the extended symmetry group of the theory on the multimomentum phase space. Such an action gave rise to the covariant momentum map, which was fundamental to recovering the admissible Cauchy data of the system when performing the space plus time decomposition of the model, specifically, we were able to recover the correct constraint structure plus the extended $t$-instantaneous Hamiltonian of the model. Furthermore, we also introduced the polysymplectic formalism for the model of our interest. Within the polysymplectic framework we obtained the polysymplectic form and the De Donder-Weyl Hamiltonian for the model, following the algorithm proposed by Kanatchikov in~\cite{kana4} to treat singular systems within the polysymplectic approach. In particular, we showed that the system is characterized by a set of the second-class constraints, which, in turn, led us to introduce a non-trivial Dirac-Poisson bracket to treat this second-class constraints as strong identities. Such a Dirac-Poisson bracket allowed us to obtain the correct De Donder-Weyl field equations for the Palatini-Cartan model for gravity encompassing the torsion free condition and the Einstein field equations on the primary constraint surface, thus proving the equivalence between the Lagrangian and the De Donder-Weyl-Hamiltonian field equations. We would like to point out here that, in order to construct the correct Dirac-Poisson bracket, we have used properties of a peculiar generalization of the Moore-Penrose inverse. The appearance of this generalized inverse within the polysymplectic formalism was justified by the non-vanishing kernel of the polysymplectic form $\eqref{pos1}$, which might not be the case in the usual Dirac-Hamiltonian approach. As discussed in~\cite{kana6}, a similar result for teleparallel gravity was obtained, but, as far as we know, this paper is the first in which a non-trivial Dirac-Poisson bracket has been explicitly constructed for the four-dimensional Palatini-Cartan model, within the polysymplectic approach for classical field theory, which is a topologically non-trivial, realistic, and dynamical description for gravity. In addition, the Dirac-Poisson bracket constructed here may be regarded  as a fundamental building block towards the quantum analysis of the Palatini-Cartan model within the so-called pre-canonical quantization approach for the De Donder-Weyl canonical theory~\cite{kana5}. Further, we have briefly discussed the conserved currents of the theory within the De Donder-Weyl Hamiltonian approach through the covariant momentum map induced by the action of the extended gauge symmetry group on the polymomentum phase space.   \\
Moreover, we introduced a slicing in the spacetime manifold, which induces a Cauchy surface foliation on it. This, in turn, allowed us to induce foliations on the different fiber bundles that describe each of the geometric frameworks presented in this paper. Thus, after performing the space plus decomposition of the affine jet bundle, we recovered the $t$-instantaneous Lagrangian function that describes the Palatini-Cartan gravity system, which in light of the $t$-instantaneous Legendre transformation enabled us to obtain the instantaneous momentum variables of the theory and also define its associated primary constraint surface. We also performed a space plus time decomposition at the multisymplectic level, where we found that the covariant momentum map associated with the extended symmetry group of the system gave rise to a momentum map on the $t$-instantaneous multimomentum phase space. In particular, we proved that the symmetries of the model are localizable symmetries and, as a consequence, the zero level set of such a covariant momentum map corresponds to the surface on the $t$-instantaneous phase space characterized by the complete set of first-class constraints of the model, which coincide with the set of first-class constraints within the instantaneous Dirac-Hamiltonian analysis of the system, as described in~\cite{nikolic1}. Lastly, after an adequate gauge fixing, we recovered the $t$-instantaneous Hamiltonian of the theory as a linear combination of the first-class constraints on the reduced phase space as discussed in~\cite{nikolic1, canepa1,canepa2}. 

Taking all these results into account, we have shown that, the  geometric-covariant Lagrangian, multisymplectic and polysymplectic formalisms for classical field theory describes the features of the Palatini-Cartan model for gravity in a geometric and consistent way. Specifically, we have described how the instantaneous constrained structure of the model, which appears in its standard instantaneous Dirac-Hamiltonian analysis, can be easily recovered using the zero level set of the covariant momentum maps associated with the symmetries of the system when we perform the space plus decomposition of its multisymplectic formulation. Moreover, based on the results of the present work, we can assure that the algorithm presented in~\cite{kana4} to study singular Lagrangian system in the framework of the polysymplectic formalism is adequate for the analysis of non-trivial gravitational models. In addition, our polysymplectic study of the Palatini-Cartan model provides a non-trivial Dirac-Poisson bracket, constructed under condition $\eqref{pa10}$, therefore, our work may shed some light on a deeper understanding of this bracket structure for gravitational models, which turns out to be fundamental to recover the correct De Donder-Hamiltonian field equations associated with  the model under consideration. Also, this Dirac-Poisson bracket may be considered as a starting point towards a quantum analysis of the Palatini-Cartan model within the precanonical quantization program, which is strongly based on the polysysmplectic formalism for classical field theory~\cite{kana5}. Certainly, there is still a lot of work to be done on non-trivial models for gravity, for example, the geometric-covariant analysis for the Asthekar's formulation of General Relativity has not yet been reported. This will be done elsewhere.

\section*{Acknowledgments}
The authors would like to acknowledge support from SNII SECIHTI-Mexico and finacial support from COPOCYT under the project 2467 HCDC/2024/SE-0416 (Convocatoria 2024-03, Fideicomiso 23871). Jasel Berra-Montiel acknowledges financial support from Marcos Moshinsky foundation.

\section*{References}

\bibliographystyle{unsrt}

\end{document}